\documentclass[aps,prl, superscriptaddress, floatfix, preprint]{revtex4-2}
\usepackage{graphicx}
\usepackage{siunitx}
\usepackage{hyperref}
\usepackage{comment}
\begin{document}

%\preprint[preprintnumbers]{}

% Alternate titles: title{Traceable randomness with a quantum advantage}
% title{Traceable randomness from quantum nonlocality}
%\title{Random numbers traceable to nonlocal correlations}
\title{Traceable random numbers from a nonlocal quantum advantage}
%\title{A third-party-verifiable source of device-independent random numbers, on demand}
%\title{Random numbers }

\author{Gautam A. Kavuri}\email{gautam.kavuri@colorado.edu}
\author{Jasper Palfree}
\author{Dileep V. Reddy}
\affiliation{Department of Physics, University of Colorado, Boulder, CO, 80309, USA}
\affiliation{Associate of the National Institute of Standards and Technology, Boulder, CO, 80305, USA}
\author{Yanbao Zhang}
\affiliation{Quantum Information Science Section, Computational Sciences and Engineering Division, Oak Ridge National Laboratory, Oak Ridge, Tennessee 37831, USA}

\author{Joshua C. Bienfang}
\affiliation{Joint Quantum Institute, National Institute of Standards and Technology and University of Maryland, 100 Bureau Drive,
Gaithersburg, Maryland 20899, USA.}
\author{Michael D. Mazurek}
\affiliation{Department of Physics, University of Colorado, Boulder, CO, 80309, USA}
\affiliation{Associate of the National Institute of Standards and Technology, Boulder, CO, 80305, USA}
\author{Mohammad A. Alhejji}
\affiliation{Center for Quantum Information and Control, University of New Mexico, Albuquerque, NM, 87131, USA}
\author{Aliza U. Siddiqui}
\affiliation{Department of Electrical, Computer, and Energy Engineering, University of Colorado, Boulder, Colorado 80309, USA}
\author{Joseph M. Cavanagh}
\altaffiliation[Current address: ]{ Pitzer Center for Theoretical Chemistry, Department of Chemistry, University of California, Berkeley, California, 94720, United States}
\author{Aagam Dalal}
\affiliation{Physical Measurement Laboratory, National Institute of Standards and Technology, Gaithersburg, MD 20899, USA.}
\author{Carlos Abell\'an}
\author{Waldimar Amaya}
\affiliation{Quside Technologies S.L., Castelldefels (Barcelona), Spain}
\author{Morgan W. Mitchell}
\affiliation{ICFO-Institut de Ciencies Fotoniques, The Barcelona Institute of Science and Technology, 08860 Castelldefels (Barcelona), Spain.}
\affiliation{ICREA - Instituci\'{o} Catalana de Recerca i Estudis Avan{\c{c}}ats, 08010 Barcelona, Spain}
\author{Katherine E. Stange}
\affiliation{Department of Mathematics, University of Colorado, Boulder, CO, 80309, USA}
\author{Paul D. Beale}
\affiliation{Department of Physics, University of Colorado, Boulder, CO, 80309, USA}
\author{Lu\'is T.A.N. Brand\~ao}
\affiliation{Strativia (Contractor Foreign Guest Researcher at NIST
Cryptographic Technology Group), Gaithersburg, MD 20899, USA}
\author{Harold Booth}
\author{Ren\'e Peralta}
\affiliation{Information Technology Laboratory, National Institute of Standards and Technology, Gaithersburg, MD 20899, USA.}
\author{Sae Woo Nam}
\affiliation{Department of Physics, University of Colorado, Boulder, CO, 80309, USA}
\affiliation{Physical Measurement Laboratory, National Institute of Standards and Technology, Boulder, CO, 80305, USA}
\author{Richard P. Mirin}
\affiliation{Physical Measurement Laboratory, National Institute of Standards and Technology, Boulder, CO, 80305, USA}
\author{Martin J. Stevens}
\affiliation{Physical Measurement Laboratory, National Institute of Standards and Technology, Boulder, CO, 80305, USA}
\author{Emanuel Knill}
\affiliation{Department of Physics, University of Colorado, Boulder, CO, 80309, USA}
\affiliation{Center for Theory of Quantum Matter, University of Colorado, Boulder, CO, 80305, USA}
\affiliation{Applied and Computational Mathematics Division, National Institute of Standards and Technology, Boulder, CO, 80305, USA}
\author{Lynden K. Shalm}\email{krister.shalm@colorado.edu}
\affiliation{Department of Physics, University of Colorado, Boulder, CO, 80309, USA}
\affiliation{Physical Measurement Laboratory, National Institute of Standards and Technology, Boulder, CO, 80305, USA}
\affiliation{Quantum Engineering Initiative, Department of Electrical, Computer, and Energy Engineering, University of Colorado, Boulder, CO, 80309}

%\thanks{}


\date{\today}

\maketitle

\textbf{The unpredictability of random numbers is fundamental to both digital security~\cite{buchmann_introduction_2004-1, eastlake_3rd_randomness_2005}  and applications that fairly distribute resources~\cite{stone_why_2007, duxbury_random_1999}.
However, existing random number generators have limitations—the generation processes cannot be fully traced, audited, and certified to be unpredictable. The algorithmic steps used in pseudorandom number generators~\cite{menezes_handbook_2018} are auditable, but they cannot guarantee that their outputs were a priori unpredictable given knowledge of the initial seed. Device-independent quantum random number generators~\cite{bierhorst_experimentally_2018, liu_device-independent_2018, colbeck_quantum_2011, pironio_random_2010} can ensure that the source of randomness was unknown beforehand, but the steps used to extract the randomness are vulnerable to tampering. Here, for the first time, we demonstrate a fully traceable random number generation protocol based on device-independent techniques. Our protocol extracts randomness from unpredictable non-local quantum correlations, and uses distributed intertwined hash chains to cryptographically trace and verify the extraction process. This protocol is at the heart of a public traceable and certifiable quantum randomness beacon that we have launched~\cite{noauthor_curby_nodate-1}. Over the first 40 days of operation, we completed the protocol 7434 out of 7454 attempts---a success rate of 99.7\%. Each time the protocol succeeded, the beacon emitted a pulse of 512 bits of traceable randomness. The bits are certified to be uniform with error times actual success probability bounded by $2^{-64}$. The generation of certifiable and traceable randomness represents one of the first public services that operates with an entanglement-derived advantage over comparable classical approaches.}

\subsection{Introduction}

Random number generation underpins cryptography and security~\cite{menezes_handbook_2018, vadhan_does_2007}, but establishing the a priori unpredictability and subsequent integrity of random numbers continues to be a major challenge~\cite{stipcevic_true_2014}. A random number generator (RNG) can be separated into two logical parts~\cite{koc_cryptographic_2009, herrero-collantes_quantum_2017}: a source of entropy and a system to harness this entropy and process it into a random bit stream. A traceable RNG makes every step in the process, from the certification of the unpredictability of the entropy source through to final extraction, auditable and verifiable. This eliminates the need to rely on extraneous assumptions of trust in the random number generation process. Traceable RNGs are of particular interest to security or high-stakes systems that rely on random numbers and must allow for audits and external verification. One such application is public randomness beacons~\cite{rabin_transaction_1983, kelsey_randomness_2019}---services that periodically broadcast public random numbers. These random numbers can be used in a wide range of applications, such as public resource lotteries~\cite{bonneau_bitcoin_2015}, financial audits, jury duty selection, choosing parameters in public cryptographic schemes~\cite{syta_scalable_2017, kelsey_randomness_2019}, and voting machine sampling~\cite{lopez_2011_2011}.

Being able to verify and audit the random numbers these beacons publish---to detect potential tampering and demonstrate a priori unpredictability---is vital to establishing public trust for their use in these applications.

Implementing a fully traceable RNG is an outstanding challenge. Algorithmic pseudo-RNGs can be audited to verify that the algorithm was executed correctly~\cite{micali_verifiable_1999, syta_scalable_2017}. However, since pseudo-RNGs are deterministic by construction, an adversary who discovers the initial inputs can perfectly predict its outputs. Hardware RNGs use a physical source of entropy that is in principle non-deterministic, but require establishing and trusting models of the components of the entropy source and the details of how the random numbers are extracted~\cite{gras_quantum_2021, abellan_generation_2015}. The reliance on device models makes it impossible to directly certify~\cite{pironio_random_2010} the a priori unpredictability of the entropy source, and tracing the outputs to the source must be predicated on trusting these unverifiable device models.

Protocols for random number generation that use such hardware RNGs are not fully traceable because these device assumptions can be undermined to allow for side channels that either leak information to an adversary~\cite{soucarros_influence_2011} or open backdoors that enables control of the output random bits~\cite{ragab_crosstalk_2021}. 

Device-independent sources of randomness~\cite{bierhorst_experimentally_2018, liu_device-independent_2018, shalm_device-independent_2021-1, li_experimental_2021} eliminate the need for detailed device models. They rely on a series of trials involving non-local measurements made on distributed entangled particles, akin to loophole-free tests of Bell's inequalities~\cite{shalm_strong_2015, giustina_significant-loophole-free_2015, hensen_loophole-free_2015,  rosenfeld_event-ready_2017, storz_loophole-free_2023}. As long as measurement basis choice and detection events at the measurement stations are space-like separated, the measurement outcomes cannot be determined a priori. 
The minimal assumptions required to certify this a priori unpredictability are the impossibility of faster-than-light signaling, that the measurement settings choices at each station are made independently of the experimental devices, that the timing and distance measurements of the stations are accurate, and that the recording devices are operating faithfully and are secure~\cite{bierhorst_experimentally_2018}. The ability to directly certify the random source without requiring a device model represents a true quantum advantage. For our traceable RNG demonstration, we employ a device-independent source of entropy that is based on measurements made on polarization-entangled photons produced via spontaneous parametric down conversion (SPDC). The mathematical certification of entropy in the outputs~\cite{zhang_certifying_2018,dupuis_entropy_2020} of this source requires a hypothesis test for probability estimation that uses accumulated products of probability estimation factors (PEFs) and can, conditional on a rejection of the null hypothesis, provide an upper bound on the probability of any allowed output conditional on the input settings choices and any other classical side information of an adversary. PEFs are functions of the basis choices and outcomes of a trial. They must, along with a number of other parameters, be fixed before performing that trial~\cite{supplementary_info}. 

The raw outcome bit string from our SPDC-based loophole-free Bell-test trials is highly biased, while a uniform bit string is desirable for most applications. Extractor functions  ~\cite{ma_postprocessing_2013, luca_trevisan_extractors_2001} can be applied to a biased random string, along with independent randomness and a certificate of entropy, to generate a shorter string that is very close to uniform. In our demonstration, we use Maurer, Portmann, and Scholtz’s implementation of the Trevisan extractor (TMPS extractor) \cite{mauerer_modular_2012} to generate a close-to-uniform output. The certificate of entropy from the entropy estimation stage can be composed with the classical-proof TMPS extractor to yield a complete protocol for the generation of uniform device-independent random bits,~\cite{supplementary_info}. While demonstrations of random number generation based on such sources have been performed before~\cite{zhang_experimental_2020, bierhorst_experimentally_2018, liu_device-independent_2018}, they fall short of achieving traceability. This is because the experiment, certification, and extraction were carried out in an opaque, offline, or one-off manner. Building a traceable RNG that performs these steps verifiably and on-the-fly is an outstanding challenge, especially if the RNG is continuously operational. 

Our goal is to develop a traceable source of certified randomness, that can also be incorporated as a service into a larger public digital ecosystem. Our approach for achieving traceability is based on the idea of distributing authority for protocol correctness and data integrity. Instead of a single party controlling the entire randomness generation process, we distribute the protocol between multiple independent parties that must work together to produce the randomness. Every action each party takes must be recorded in a tamper-resistant manner that can be independently verified or audited. 

The solution, a protocol we call Twine, is based around the concept of intertwining different hash chains to form a hash graph~\cite{lamport_password_1981, thulasiraman_graphs_2011}. A hash chain, sometimes known as a block chain, is a cryptographically secure time-ordered data structure. A new block is added to the chain by hashing~\cite{dworkin_sha-3_2015} input data, such as a record of a stage of  the randomness generation process that one party wishes to store, with the hash of the previous block. This creates an ordered chain of data. Attempting to change a block of data without detection would require rewriting the entire chain after that point. To prevent this kind of tampering with all the blocks on a single chain, the Twine protocol allows a block to include the hashes of blocks from multiple chains operated by different parties, creating a directed acyclic graph (DAG)~\cite{cormen_introduction_2009}. If a party tries to tamper with their published records, it can be detected by other parties, since the hashes recorded on their chains will no longer be consistent. In other words, for a bad actor to go undetected, they need to surreptitiously rewrite the history of not only their own hash chain, but also of everyone else that is connected. As the number of independent parties in the network grows, such an action becomes increasingly difficult.

We use Twine to create a traceable time-ordered cryptographic contract between three parties, each responsible for a part of the device-independent randomness generation process. These parties are a part of the University of Colorado at Boulder (CU) randomness beacon network we call CURBy. The first party is the National Institute of Standards and Technology (NIST), which performs the Bell test that serves as the raw source of entropy for the protocol. The second party is at CU which runs a quantum randomness (CURBy-Q) process that sets up the hypothesis test, analyzes the data, and runs the extractor. The third party is an external randomness service, the Distributed Randomness Beacon Daemon (DRAND)~\cite{noauthor_dranddrand_2024}, which provides an independent seed to the extractor. The steps of the protocol, as displayed in Fig.~\ref{protocol figure}, are time ordered and auditable. This means the hypothesis used in the data analysis is published and committed to before the Bell data for the device-independent source is taken, and all inputs for the extraction (raw data, certificate, and seed) are committed to before knowledge of the seed (released by DRAND) could be known. In this way, no single party has complete control over the output random bits, and the time-ordering, freshness, and integrity of the data can be verified and audited. 

The larger CURBy Network also involves other parties (more details in the supplemental material~\cite{supplementary_info}). This includes a second traditional randomness beacon, named CURBy-RNG, that is closely based on the NIST standard for random beacons~\cite{kelsey_randomness_2019}, and emits 512 bits of randomness every \qty{60}{\second}. The output of the traceable quantum randomness beacon acts as a cryptographic salt~\cite{grassi_digital_2017} for the CURBy-RNG beacon via hash propagation provided by the Twine protocol. Finally, an independent time-stamping chain is used to provide certified timestamping of the network using a traditional certified timestamp service. Together, these chains form a robust source of decentralized randomness that can be incorporated into emerging decentralized internet protocols~\cite{benet_ipfs_2014} through an accessible application programming interface~\cite{noauthor_curby_nodate-1}.

\subsection{Protocol details}
\begin{figure} [b!]
  \includegraphics[width=0.7\linewidth]{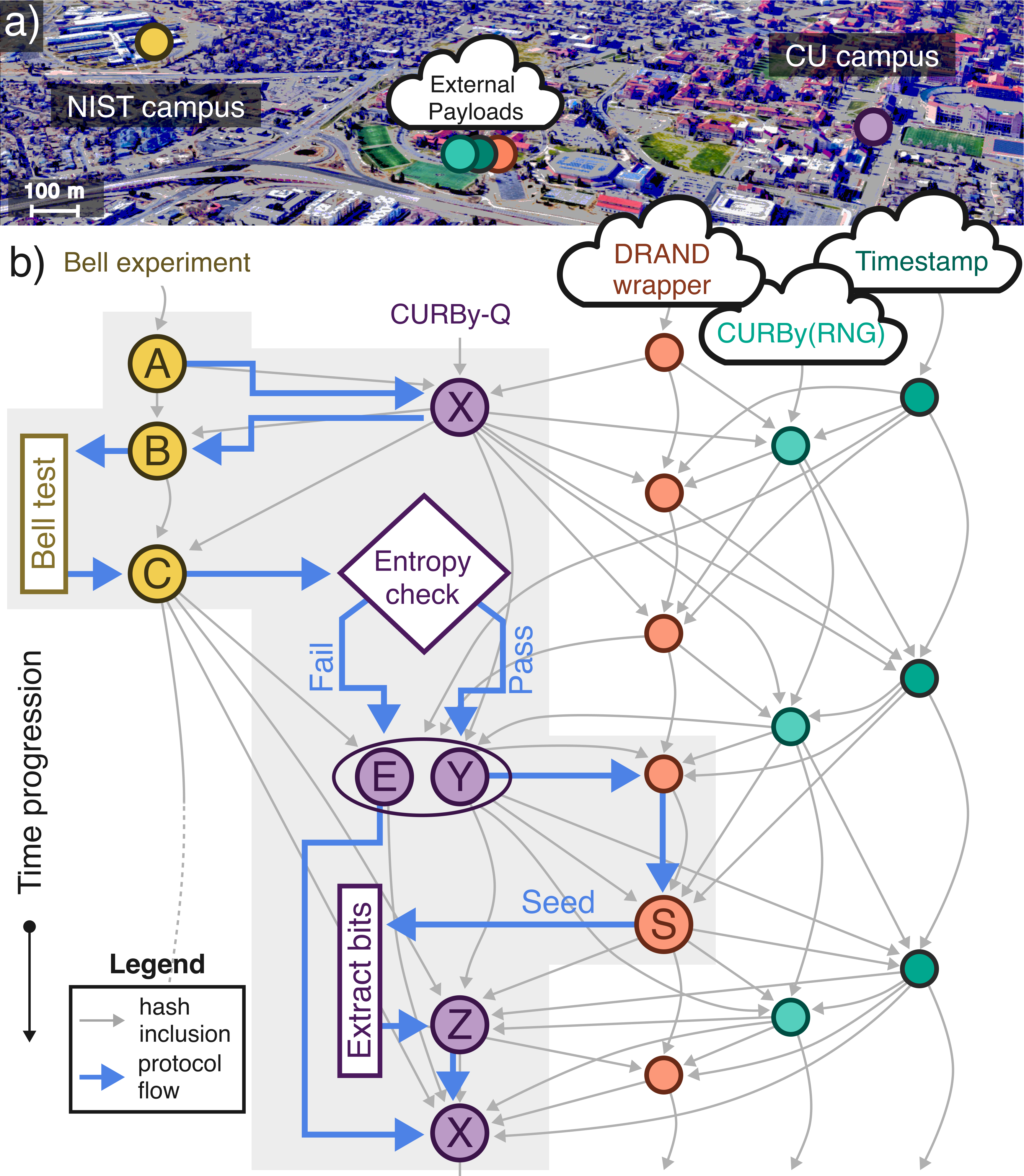}
\caption{(a) Aerial photograph of Boulder, Colorado indicating the two geographically separate parties and external parties involved in the protocol. The Bell test experiment is run at the NIST campus. The certification and extraction of randomness is performed at the University of Colorado (CU). Three hash chains include external payloads and are indicated under the cloud.   (b) Diagram of the verifiable device-independent RNG protocol followed in this demonstration, and the geographically separate parties involved. The schematic follows one instance of the protocol from status update to publishing of the result, with published pulses represented as colored-in circles. Differently colored circles form distinct hash chains that track independent processes. }  \label{protocol figure}
\end{figure}

The traceable RNG demonstration involves five primary hash chains as shown in Fig.~\ref{protocol figure}. Additional hash chains coordinate third-party interactions that further decentralize the trust from the authority in control of the physical device-independent source. Four of the primary hash chains are processed on machines at CU.
 
The fifth hash chain (Bell test experiment) tracks the generation and collation of raw Bell trial data from the device-independent source, and is run on computers at NIST. The protocol begins with a request for raw Bell trial data, which is registered on the CURBy-Q chain. This request includes precommitment of a number of parameters that are used in a hypothesis test to certify entropy in the Bell trial data, and subsequently extract random numbers from it. Upon receipt of a request (pulse B in Fig.~\ref{protocol figure}), computers at NIST prepare the experiment to start performing loophole-free Bell trials. 

A Bell trial at NIST begins with the generation of a pair of polarization-entangled photons. We start with a $\qty{774.3(2)}{\nano \meter}$ (see methods~\cite{methods} for a note on how we report uncertainties) wavelength pulsed laser with a pulse repetition rate of $\qty{80.00\pm0.01}{MHz}$ at the source station, which is incident on a periodically-poled potassium titanyl phosphate (ppKTP) crystal inside a Mach-Zehnder interferometer, see Fig.~\ref{experimental figure}. This results in probabilistic generation of polarization-entangled pairs of photons close to \qty{1550}{nm}. Given that a single downconversion event occurs, the state prepared is nominally $0.383 |\mathrm{HH}\rangle + 0.924
|\mathrm{VV}\rangle$, where the H and V represent horizontal and vertical polarizations, respectively. The photons are then distributed via optical fiber to two remote stations (Alice and Bob) separated by $\approx \qty{110}{\meter}$ (see Fig.~\ref{experimental figure}). While the photons are in transit, hardware RNGs at Alice and Bob each make random binary settings choices that are fed to Pockels cells that rapidly set the polarization basis in which the photons will be measured. The distributed photons then arrive at the stations and are measured in one of two bases---Alice chooses between $a = \ang{6.7}$ and $a^{\prime} = \ang{-29.26}$ and Bob chooses between $b = \ang{-6.7}$ and $b^{\prime} = \ang{29.26}$, where the angles represent rotations of a linear polarization state relative to a horizontally oriented polarizer. The photons are detected with superconducting nanowire single-photon detectors (SNSPDs) with $>97\%$ system detection efficiency~\cite{reddy_superconducting_2020}. The end-to-end efficiency, from the generation of photon pairs to their detection, is $\approx 81\%$. A trial is considered complete when the outcomes from the detectors are recorded on timetaggers. To enforce non-locality in the Bell test, the outcomes must be recorded at Alice (Bob) before hypothetical light-spheres containing settings-choice information propagating from Bob (Alice) are able to reach the timetaggers. We assume that the outcome (a classical electrical signal), once recorded on the trusted timetaggers, is not altered. For every trial, we check that such non-locality is enforced~\cite{supplementary_info}, and declare invalid any data that contains trials that do not satisfy the timing constraints. Completing 15 million such trials takes $\approx \qty{60}{\second}$, after which the data is packaged and passed privately to computers at the University of Colorado (CU). A hash (checksum) of the data is posted publicly to allow for auditing.

Computers at CU attempt to certify 820 bits of min-entropy in the outputs of the completed run of the Bell test experiment~\cite{zhang_certifying_2018} (this is adjusted to the smoothness error used in the demonstration,~\cite{methods}). If successful, the computers extract 512 uniform bits from the output string. The TMPS extractor requires independent uniform seed bits to perform the extraction. To provide these bits after every successful certification, the computers at CU publicly commit to using, and then await, the next random pulse from the DRAND beacon (distributed randomness beacon)~\cite{syta_scalable_2017, noauthor_dranddrand_2024}, which is algorithmically expanded~\cite{methods}. This interaction is mediated via the DRAND wrapper chain, see Fig.~\ref{protocol figure}.

\subsection{Summary of results}

We demonstrate our protocol through an aperiodic, traceable randomness beacon~\cite{rabin_transaction_1983} that publishes pulses (data structures containing random numbers and meta-information~\cite{supplementary_info}) of 512 certified random bits with a protocol soundness error $\epsilon = 2^{-64}$ (the soundess error is defined as the product of uniformity error and success probability~\cite{supplementary_info, methods}). Because the inputs, outputs, and algorithms used at each stage of the protocol are publicly known or recorded on the DAG, this beacon is, to our knowledge, the first random number generator that can not only certify the quality of the randomness but also enables traceability by providing a complete audit trail that verifies the freshness, time ordering, and integrity of each step of the randomness generation process. This is an early example of a publicly available service that operates with a provable quantum advantage~\cite{a_g_j_macfarlane_quantum_2003}. 

A key requirement for a random number service is high availability and uptime. Over the initial 40 days of operation of our traceable beacon service, CURBy-Q, the randomness generation protocol was run a total of 7454 times and succeeded in generating traceable certified randomness 7434 times (a success rate of $99.7\%$). Most of the failures came from software errors in our timetaggers during data taking. On average, we were able to produce 186 randomness pulses per day, with $95\%$ of the pulses during normal operation taking between \qty{207.1 \pm 0.1}{\second} and \qty{327.7 \pm 0.1}{\second} to produce, as shown in Fig.~\ref{latency figure}. This time was dominated by the classical processing steps required in the protocol. The Bell test is only operational for approximately 17 hours a day due to the need to recycle the cryostats used for the superconducting detectors. Once a day, when the cryostats are warm when a request is made, the requests take approximately 7 hours to fulfill. Data on the continued operation of experiment beyond the first 40 days is available online~\cite{noauthor_curby_nodate-1}.

The TMPS extractor has some overhead associated with its operation conditional on the allowed error. Consequently, to output 512 bits of randomness certified uniform with a soundness error of $2^{-64}$, the raw outputs from the Bell test need to contain at least 820 bits of min-entropy (adjusted to the smoothness error used). Every time the protocol is run, Bell test data is acquired for \qty{60}{\second}. The running accumulated entropy for every pulse generated over the 40 days is shown in Fig.~\ref{running entropy figure}. For $95\%$ of the pulses served, it took between \qty{6.5 \pm 0.6}{\second} and \qty{18.4 \pm 0.6}{\second} to cross the 820 bit entropy threshold needed for the protocol to succeed (mean time is \qty{10.844 \pm 0.007}{\second}). This represents a nearly 15 times speedup over our past device-independent RNG hardware demonstrations~\cite{zhang_experimental_2020}, which is paired with higher stability and uptime. The increased performance is due to improvements in the overall system efficiency from $76\%$~\cite{zhang_experimental_2020} to $\approx 81\%$ for the Bell test setup. Small daily changes in the alignment of the system lead to different effective slopes (rates) of accumulated entropy. Increasing the entropy threshold would allow for far more bits to be extracted per data set, but at the cost of an increased chance of the protocol failing.

\subsection{Conclusion}

The traceability of our randomness beacon comes in part from the distributed nature of the protocol. Multiple independent authorities, who may not trust one another, work together to create the random outputs in a verifiable manner. Expanding this idea, it is possible to create many independent randomness beacons that all intertwine with one another to produce networked randomness to which each beacon contributes, but no one beacon controls. The randomness beacons may run different protocols and specifications, but by cross feeding their results into the larger network hash graph the security and trustworthiness of the entire system increases. 

We have developed an application programming interface (API) that makes it simple to use the randomness generated by this network, to audit and verify the results of individual beacons or other computable contracts created on the network, and to launch other independent randomness beacons or other data services that automatically intertwine with the larger Twine hash graph. Twine also integrates into emerging decentralized web technologies, such as peer-to-peer distributed file storage systems~\cite{benet_ipfs_2014}, based on hash graphs and blockchains. In this context, CURBy-Q regularly injects an external source of certified randomness that acts as a kind of cryptographic salt whose influence spreads throughout the network and helps the entire system become more secure. To the best of our knowledge, our quantum randomness beacon is the first quantum 2.0~\cite{a_g_j_macfarlane_quantum_2003} service that can provide a true quantum advantage to help improve the security and trust of next-generation web technologies.

Intertwined hash graphs can be used to add trust, security, and traceability to other classical and quantum protocols through computable contracts. They are particularly well suited to applications that require making precommitments, such as public scientific hypothesis tests and registered reports~\cite{chambers_past_2021}. Such intertwined operation can be used to increase the trustworthiness of classical-verifier-based protocols for remote quantum state preparation~\cite{gheorghiu_computationally-secure_2019, cojocaru_qfactory_2019}, blind and verifiable quantum computation~\cite{gheorghiu_computationally-secure_2019, mahadev_classical_2023}, and quantumness proofs~\cite{zhu_interactive_2023, brakerski_simpler_2020}. It can also be used to ensure that quantum computing benchmarking tests are run fairly by providing a time-sensitive audit trail. Beyond the creation of a quantum randomness service operating with a quantum advantage, our work is an important step towards the symbiotic integration of entanglement-assisted quantum communications and security protocols with emerging internet technologies.

\clearpage

\section{Main Text Figures}

\addtocounter{figure}{-1}
\begin{figure} [t!]
  \caption{ Hash connections, represented by gray arrows, serve to provide verifiable time ordering and data integrity. After the entropy check, only one of the ``E" or ``Y" pulses is published, depending on the raw data passing or failing the check. Pulse labels are A: Status; B: Request queued; C: Bell test complete; E: Error; S: External seed; X: Data request; Y: Precommit; Z: Output randomness. See supplementary material~\cite{supplementary_info} for more information on chain construction. Imagery \copyright 2022 Google, Imagery \copyright 2022 CNES/Airbus, Maxar technologies, US Geological Survey USDA/FPAC/GEO, Map data \copyright 2022}
\end{figure}

\begin{figure*}
\includegraphics[width=\linewidth]{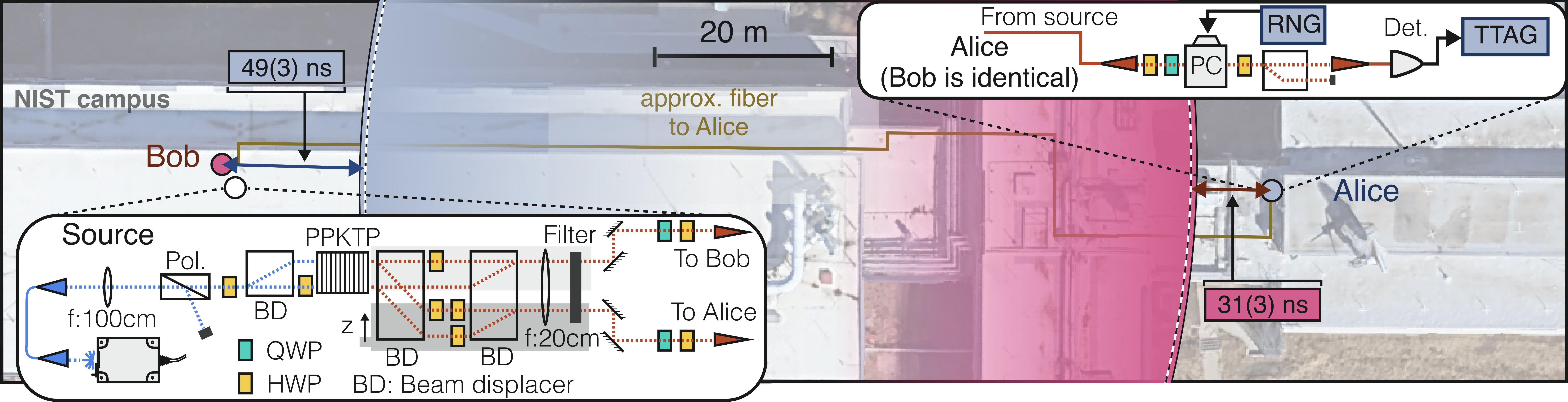}
\caption{Satellite overview of the experiment (at the NIST campus), with the maximum spatial extent of an information lightcone from the measurement choice at the Alice (Bob) station at the moment when the corresponding measurement is complete at the Bob (Alice) station indicated by the blue (pink) arc. Lightcones are constructed based on responses to data requests from the first 40 days. The white dashed lines are the mean extent, and the colored in lightcones represent the region of maximum extent after taking standard uncertainty into account. 
Insets are schematics of the entangled-pair generation source and measurement station used for performing space-like separated Bell trials. Polarization-entangled pairs of photons are generated in a periodically poled potassium titanyl phosphate (ppKTP) crystal with a set of waveplates and beam displacers. The photons are coupled into single-mode fibers and delivered to two remote measurement stations. At the measurement stations, a binary choice of measurement is made with a hardware random number generator (RNG), and fed to a Pockels cell (PC), which changes the polarization of the photons based on the choice. Finally, the results of binary-outcome measurements made on the photons are recorded on timetaggers (TTAG). Imagery \copyright 2023 Airbus, CNES/Airbus, Maxar Technologies, U.S. Geological Survey, Map Data \copyright 2023 Google  \label{experimental figure}}
\end{figure*}

\begin{figure}
\includegraphics[width=\linewidth]{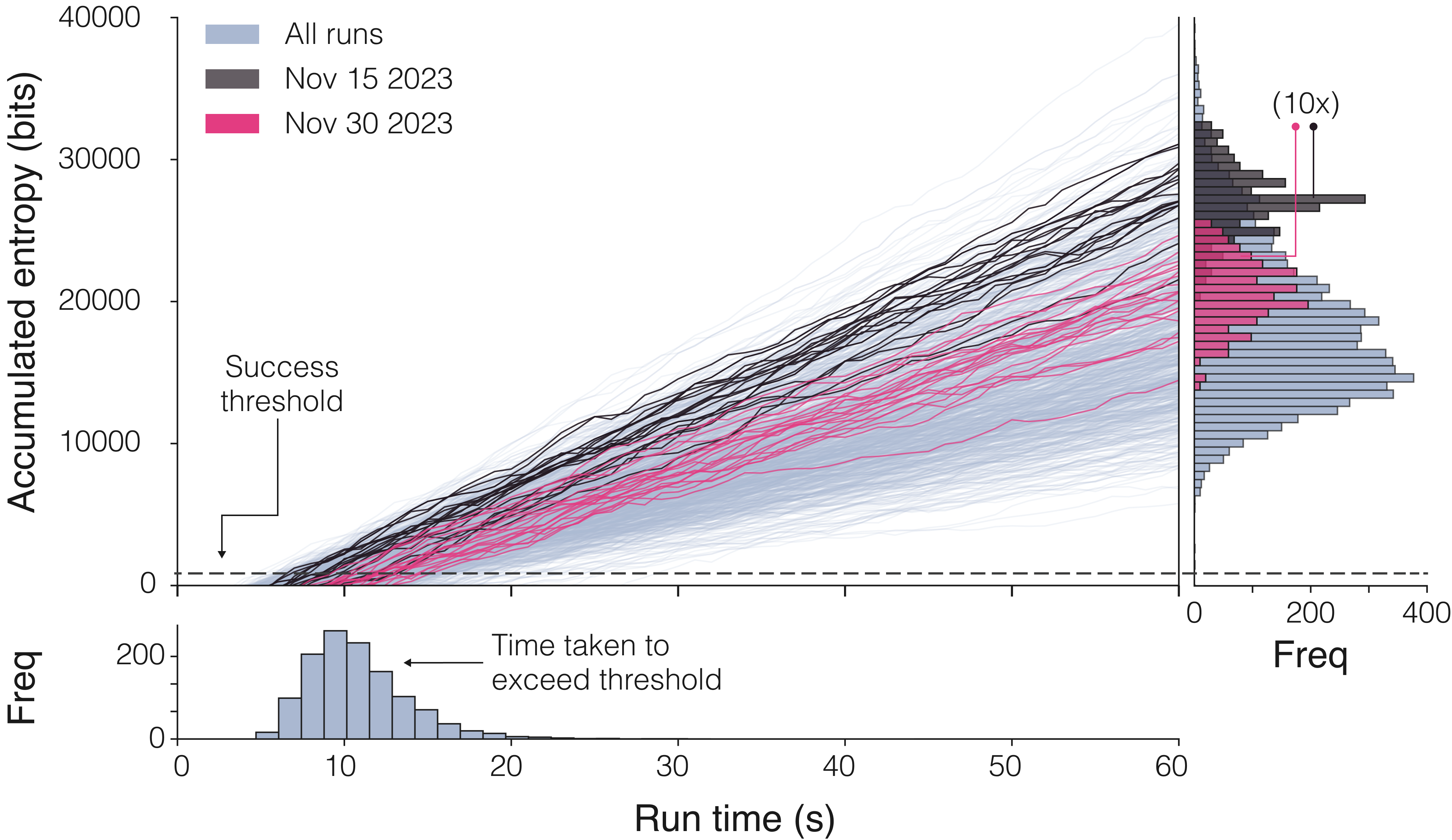}
\caption{Plot of the running entropy estimate of 1000 sets of Bell trial data---sampled uniformly from 7434 successful responses to 7454 data requests made over 40 days---as a function of truncation time, in blue. Also plotted are two sets of 20 running entropy estimates each uniformly sampled from responses on two chosen days (November 15, 2023 and November 30, 2023) in black and magenta. Below the plot is a marginal histogram of the time required for all the 7434 data sets to cross the threshold of 820 bits (horizontal dotted line on main plot). On the right is a marginal histogram of the final entropy of all data sets in blue, and the final entropy of random output pulses served on the two specific days in black and magenta, during which 182 and 181 random number pulses were respectively served.   \label{running entropy figure}}
\end{figure}

\begin{figure}
\includegraphics[width=\linewidth]{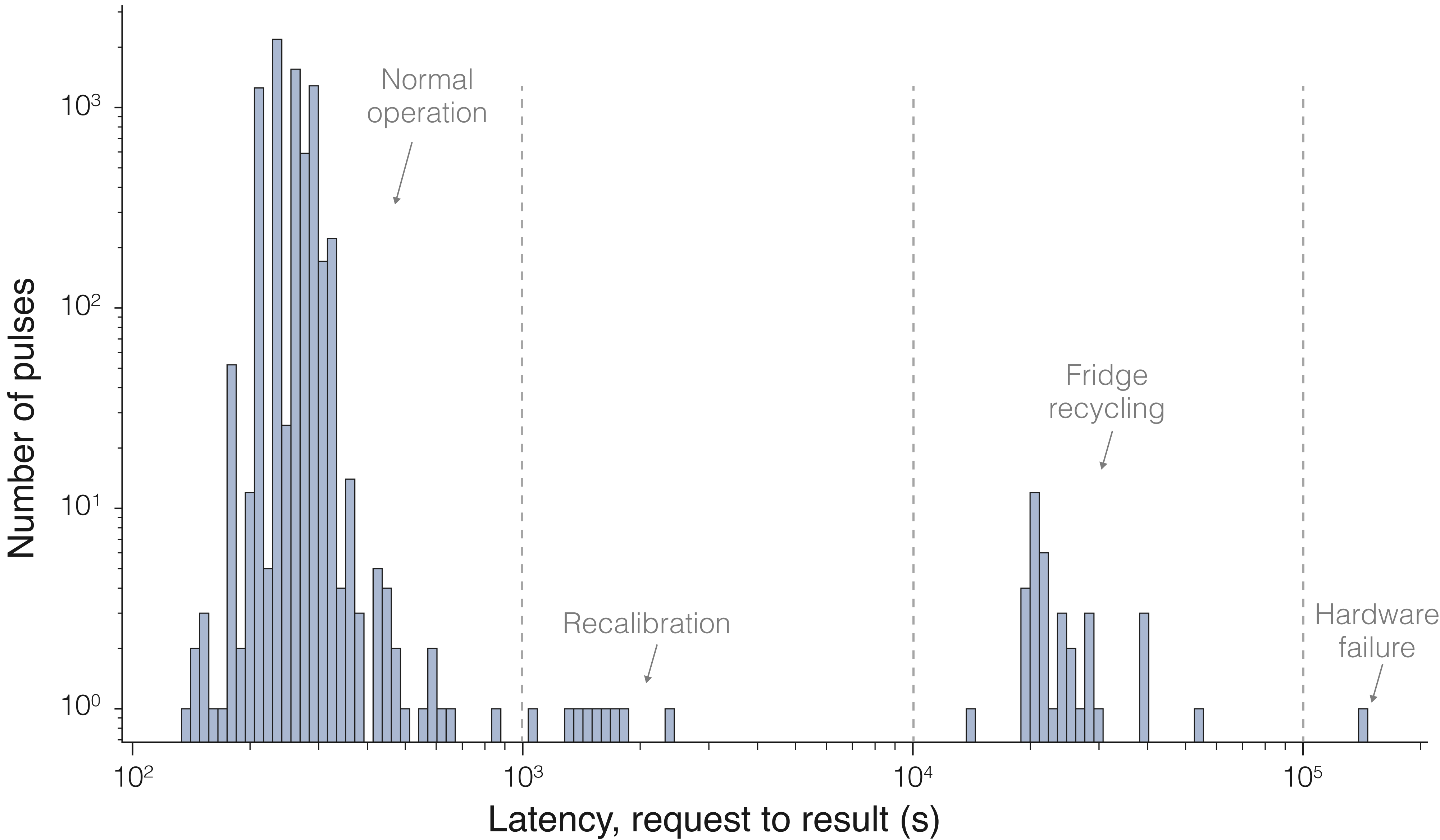}%
\caption{A histogram of the latency (in seconds) from a data request to publishing of the result for 7454 data requests over 40 days. The latencies can be divided into four geometric regimes (vertical dashed lines) depending on the status of the experiment when the request was made. Each day, the cryogenic refrigerators housing our superconducting detectors need to recycle. A total of 40 requests for randomness were made during this recharging cycle. For these requests, the experiment waits for the fridges to recycle before collecting Bell data. This resulted in total latencies of more than \qty{10000}{\second}. On one day, a hardware failure caused a fridge to take about 40 hours to recycle, resulting in the outlier. On occasion, the experiment runs an automated recalibration. This recalibration was running during 9 of the requests, leading to a longer protocol completion time. The number of pulses served during the demonstration in these regimes are: Normal operation, 7406; Recalibration, 9; Fridge recycling, 38; Hardware failure, 1. Roughly $95\%$ of the pulses during normal operation took between \qty{207.1 \pm 0.1}{\second} and \qty{327.7 \pm 0.1}{\second} to produce. \label{latency figure}}
\end{figure}

\clearpage

\section{Methods}
\subsection{Extended protocol details}
The protocol followed for the end-to-end random number generation follows a request response model, with raw data continually requested from the NIST experiment, and subsequently processed independently by other parties. We detail the protocol below. 

Prior to the collection of data from the NIST experiment setup at the beginning of the demonstration and before every randomness generating round, we precommit to probability estimation factors (PEFs), the power $\beta$~\cite{zhang_certifying_2018}, the desired length of the uniform random bit string, the stopping criterion for the number of trials, error bounds, the maximum allowed adversarial bias in the random inputs, and the number of seed random bits needed to perform the randomness extraction. This precommitment is published on the CURBy-Q hash chain. The precommitted PEFs and $\beta$ above are decided upon by using published, prior data from the NIST experiment and enforcing the allowed error ($\epsilon = 2^{-64}$) and bit requirements (\qty{512}{bits}) of a standardized request. During the course of the demonstration described here, these were automatically updated about once each day.

This precommitment also serves as a request for loophole-free Bell trial data and marks the start of a round of randomness generation. Upon seeing a new request, computers and timetaggers at NIST perform loophole-free Bell trials for $\approx$ \qty{60}{\second}  (based on prior testing this integration time, corresponding to 15 million trials was more than sufficient to meet the standard request; see Fig.~\ref{running entropy figure}). The Bell trials are performed on two separated untrusted devices (conventionally called Alice and Bob). Certification of entropy in the outputs of the devices is conditional on space-like separation, or no-signaling, between the measurements at the untrusted devices. 

A certificate of space-like separation is passed to computers managing the CURBy-Q chain along with every chunk of raw data. This certificate is based on trusted measurements of a few electrical latencies and physical distances that are assumed to remain constant, along with real-time measurements of other latencies~\cite{supplementary_info}. 

 The ability to certify randomness in the output string comes from making measurements on an appropriate entangled state, such that deterministic strategies for generating the output string can be ruled out with sufficiently high confidence using the precommitted PEFs. In particular, the device-independent RNG protocol starts with two RNGs at physically separated (by $\approx \qty{110}{m}$) stations Alice and Bob making random independent choices that are the logical inputs $Z_i=X_iY_i \mid X_i,Y_i \in \{0,1\}$ to untrusted devices at the stations. These inputs are used to decide the projective measurement basis for the photons as discussed in the main text. For every input bit, the untrusted devices generate a pair of outputs $C_i= A_iB_i \mid A_i,B_i \in \{0,1\}$ within about \qty{260}{\nano \second}. Each of these outputs is the outcome of a projective measurement on one half of a (probabilistically generated) polarization-entangled pair that is distributed to the remote station from a central entanglement-generating station, as described in the main text. Timetaggers at each remote station record when these outputs are produced. This set of inputs and outputs forms a single trial. 

We define the start of a Bell trial at each station to be the earliest possible time that any information about the random bit choice the hardware RNGs make could be leaked to the environment. We estimate this time using a physical model of the hardware RNGs~\cite{abellan_generation_2015}, which is estimated relative to when a random bit is output from the RNG. The latency of output of the RNGs with respect to a trial marker---which is recorded on the timetaggers for every trial performed---is measured electronically, and we trust that this remains the same throughout the experiment. The end of a trial is when the window for allowed detections closes, based on the trial marker and a fixed detection window decided on before the start of the demonstration.

In this demonstration, we perform about \qty{250000}{} trials every second. After $\approx \qty{60}{\second}$, of data acquisition (15 million trials), data from the two remote stations is lined up based on electrical reference signals recorded at the two remote timetaggers, compressed and finally passed to computers at CU running CURBy-Q.

A hash (checksum) of the trial data (inputs $\mathbf{Z} = (Z_i)_{i=1}^{n}$  and outputs $\mathbf{C} = (C_i)_{i=1}^{n}$) is posted on the Bell experiment hash chain and the data is privately passed to the computers at the University of Colorado. Computers at the University of Colorado try to certify 820 bits of $\epsilon_{h}$-smooth min-entropy in the outputs of the Bell test experiment conditional on the settings and any classical side information~\cite{zhang_certifying_2018}, with $\epsilon_h = 0.8 \times 2^{-64}$. If successful, the computers then extract 512 uniform bits from the output string using the TMPS extractor~\cite{mauerer_modular_2012}, with extractor error $\epsilon_x = 0.2\times2^{-64}$.
The seed randomness for the extractor is obtained from expanding the 512 bits from the next DRAND beacon (distributed randomness beacon)~\cite{syta_scalable_2017, noauthor_dranddrand_2024} pulse using the SHAKE256 hash algorithm~\cite{dworkin_sha-3_2015}---an extendable output function of the SHA3 family of hash functions---to about a quarter of a million bits (exact length is committed to in advance~\cite{supplementary_info}). These seed bits are then provided along with the outputs of the Bell trials to the extractor, which outputs 512 bits certified to be uniform with soundness error $\epsilon= \epsilon_h + \epsilon_x =  2^{-64}$~\cite{supplementary_info}. For an intuitive understanding of the soundness error (which is a product of the protocol success probability and the uniformity error of the output conditioned on success), consider that even if the success probability is as small as $2^{-32}$, the uniformity error on the ouput conditional on success is at most $2^{-32}$. Different combinations of success probability and uniformity error that multiply to $2^{-64}$ are equally valid with the soundness error we employ here. The extracted bits are published on CURBy-Q as a result pulse~\cite{noauthor_curby_nodate-1}.

The experiment is automated to allow for continuous operation, and is ready to respond to requests for $\approx 17$ hours a day. Requests are continually made by the University of Colorado one minute after processing of the previous request. After initial testing, the service went live on October 26, 2023 (UTC), and we report here on operation up to December 5, 2023 (UTC), when a power outage caused an interruption in service. A full, up-to-date record of the experiment can be found at~\cite{noauthor_curby_nodate-1}. Our automated realignment and recalibration was sufficient to allow for the observed high success rate ($>99.7\%$) over this extended period with minimal manual intervention. Manual intervention was needed about once a week to restart services that were interrupted and whose restart was not automated for technical reasons, and in order to set the phase on the Mach-Zehnder interferometer at the polarization entanglement source, which slowly drifted outside the acceptable range over the course of a few weeks.

\subsection{Note on uncertainties}
All of the uncertainties reported in the main and supplementary texts are standard uncertainties (1-sigma). When we report a quantity with uncertainty as (for example) $\qty{774.3(2)}{\nano \meter}$, the number in parentheses is the numerical value of the combined standard uncertainty referred to the corresponding last digits of the quoted result. Data for the latencies in Fig.~\ref{latency figure} is based on computer logs that include the system time at the start and end of the protocol. We take the standard uncertainty of the system times to be $\qty{100}{\milli \second}$. The computers periodically poll NTP time servers to synchronize their local time to internet time, and any drifts of the local clock between polling periods can cause errors in the recorded timestamps. This $\qty{100}{\milli \second}$ uncertainty is directly reflected in the uncertainty estimate for the time bounds reported in the main text.  

In Fig.~\ref{running entropy figure}, every line is formed from 60 individual data points, which correspond to the accumulated entropy after $n$ seconds, with $n = 1, 2, \ldots ,60$. The time when this line crosses over the entropy threshold is estimated by a linear interpolation between the two contiguous points that straddle the threshold. The probability distributions that characterize the uncertainty associated with such an interpolation depend on the interpolated value, and are inconvenient to model exactly. Instead, we model the uncertainty in the estimates $x_i$ as a uniform distribution with limits $[x_i -1, x_i + 1]$. This always represents a (over) full coverage of all possible values that the estimate could take. The standard deviation this distribution is $0.58$. We report this for the $95\%$ interval limits (\qty{6.5 \pm 0.6}{\second} and \qty{18.4 \pm 0.6}{\second}), and propagate these uncertainties under the independent and identically distributed assumption to the reported mean (\qty{10.844 \pm 0.007}{\second}).

\subsection{Data availability}
Data from the beacon is public, and available at random.colorado.edu. Any other data that support the plots within this paper and other findings of this study are available from the corresponding authors upon reasonable request.

\subsection{Code availability}
The code used to run the beacon is publicly available at https://github.com/buff-beacon-project \cite{noauthor_buff_nodate}. The code that produces the results and figures presented in this work is available from the corresponding authors upon reasonable request.

\section{Acknowledgments}
This work includes contributions of the National Institute of Standards and Technology, which are not subject to US copyright. The use of trade names does not imply endorsement by the US government. The work is supported by the National Science Foundation RAISE-TAQS(award 1840223), the University of Colorado at Boulder through the ``QuEST Seed Award: A Quantum Randomness Beacon'', the Colorado Office of Economic Impact (Award No. DO 2023-0335), in part by the European Union ``NextGenerationEU/PRTR.'' Spanish Ministry of Science MCIN: project SAPONARIA (PID2021-123813NB-I00) and ``Severo Ochoa'' Center of Excellence CEX2019-000910-S. Generalitat de Catalunya through the CERCA program and grant No. 2021 SGR 01453;   Fundaci\'{o} Privada Cellex; Fundaci\'{o} Mir-Puig. This work was performed in part at Oak Ridge National Laboratory, operated by UT-Battelle for the U.S. Department of energy under Contract No. DE-AC05-00OR22725. The authors would like to thank Scott Glancy for useful discussions about the project. 

\section{Author Contributions}
G.A.K. built and performed the experiment with assistance from L.K.S., M.M., and M.J.S., and collected and analyzed data. J.P. and L.K.S. developed the Twine protocol with inputs from L.T.B, H.B., and R.P., and implemented it with assistance from J.C. and A.D. D.V.R. provided the high-efficiency detectors. Y.Z., M.A.A., A.U.S., L.K.S., G.A.K, and E.K. participated in the data analysis. J.B. provided electronics and hardware RNGs. C.A., W.A., M.W.M. provided hardware RNGs. P.B. and J.P. developed a software RNG used in CURBy. J.P., K.E.S., L.K.S., and P.B. developed the hardware and software at CU to run the CURBy network. L.K.S., S.W.N. and R.P.M. supervised the project. G.A.K. led the writing of the manuscript with all authors contributing. 

\bibliography{beacon_paper_references}

\end{document}

%\preprint{}


\title{Supplemental Material: Traceable random numbers from a nonlocal quantum advantage}

\author{Gautam A. Kavuri}
\author{Jasper Palfree}
\author{Dileep V. Reddy}
\affiliation{Department of Physics, University of Colorado, Boulder, CO, 80309, USA}
\affiliation{Associate of the National Institute of Standards and Technology, Boulder, CO, 80305, USA}
\author{Yanbao Zhang}
\affiliation{Quantum Information Science Section, Computational Sciences and Engineering Division, Oak Ridge National Laboratory, Oak Ridge, Tennessee 37831, USA}

\author{Joshua C. Bienfang}
\affiliation{Joint Quantum Institute, National Institute of Standards and Technology and University of Maryland, 100 Bureau Drive,
Gaithersburg, Maryland 20899, USA.}
\author{Michael D. Mazurek}
\affiliation{Department of Physics, University of Colorado, Boulder, CO, 80309, USA}
\affiliation{Associate of the National Institute of Standards and Technology, Boulder, CO, 80305, USA}
\author{Mohammad A. Alhejji}
\affiliation{Center for Quantum Information and Control, University of New Mexico, Albuquerque, NM, 87131, USA}
\author{Aliza U. Siddiqui}
\affiliation{Department of Electrical, Computer, and Energy Engineering, University of Colorado, Boulder, Colorado 80309, USA}
\author{Joseph M. Cavanagh}
\altaffiliation[Current address: ]{ Pitzer Center for Theoretical Chemistry, Department of Chemistry, University of California, Berkeley, California, 94720, United States}
\author{Aagam Dalal}
\affiliation{Physical Measurement Laboratory, National Institute of Standards and Technology, Gaithersburg, MD 20899, USA.}
\author{Carlos Abell\'an}
\author{Waldimar Amaya}
\affiliation{Quside Technologies S.L., Castelldefels (Barcelona), Spain}
\author{Morgan W. Mitchell}
\affiliation{ICFO-Institut de Ciencies Fotoniques, The Barcelona Institute of Science and Technology, 08860 Castelldefels (Barcelona), Spain.}
\affiliation{ICREA - Instituci\'{o} Catalana de Recerca i Estudis Avan{\c{c}}ats, 08010 Barcelona, Spain}
\author{Katherine E. Stange}
\affiliation{Department of Mathematics, University of Colorado, Boulder, CO, 80309, USA}
\author{Paul D. Beale}
\affiliation{Department of Physics, University of Colorado, Boulder, CO, 80309, USA}
\author{Lu\'is T.A.N. Brand\~ao}
\affiliation{Strativia (Contractor Foreign Guest Researcher at NIST
Cryptographic Technology Group), Gaithersburg, MD 20899, USA}
\author{Harold Booth}
\author{Ren\'e Peralta}
\affiliation{Information Technology Laboratory, National Institute of Standards and Technology, Gaithersburg, MD 20899, USA.}
\author{Sae Woo Nam}
\affiliation{Department of Physics, University of Colorado, Boulder, CO, 80309, USA}
\affiliation{Physical Measurement Laboratory, National Institute of Standards and Technology, Boulder, CO, 80305, USA}
\author{Richard P. Mirin}
\affiliation{Physical Measurement Laboratory, National Institute of Standards and Technology, Boulder, CO, 80305, USA}
\author{Martin J. Stevens}
\affiliation{Physical Measurement Laboratory, National Institute of Standards and Technology, Boulder, CO, 80305, USA}
\author{Emanuel Knill}
\affiliation{Department of Physics, University of Colorado, Boulder, CO, 80309, USA}
\affiliation{Center for Theory of Quantum Matter, University of Colorado, Boulder, CO, 80305, USA}
\affiliation{Applied and Computational Mathematics Division, National Institute of Standards and Technology, Boulder, CO, 80305, USA}
\author{Lynden K. Shalm}
\affiliation{Department of Physics, University of Colorado, Boulder, CO, 80309, USA}
\affiliation{Physical Measurement Laboratory, National Institute of Standards and Technology, Boulder, CO, 80305, USA}
\affiliation{Quantum Engineering Initiative, Department of Electrical, Computer, and Energy Engineering, University of Colorado, Boulder, CO, 80309}

%\thanks{}


\date{\today}

\begin{abstract}
\end{abstract}

%\keywords{}

\maketitle
\tableofcontents

% References should be done using the~\cite,~\ref, and \label commands
\section{Measurements for spacelike separation}
In a device-independent random number generator, it is important to ensure that the two devices on which the Bell test is performed cannot communicate with each other during the course of a trial. This translates to a set of constraints of spacelike separation for inputs provided to and outputs recorded from the untrusted devices. To verify that the relevant events satisfy these constraints of spacelike separation we use a combination of measurements on the trusted and untrusted devices in our experiment. We make the assumption that the timing and distance measurements on the trusted devices stay constant through the course of the experiment. Timing measurements on the untrusted devices are made continuously during data taking using trusted timetagger. Any data collected from the untrusted devices that does not satisfy the spacelike constraints is considered invalid, and an error signal is returned. 
\subsection{Measurements trusted to remain constant}
In our experiment, we trust that the following are operating correctly and are not compromised though the course of the demonstration:

\begin{itemize}
    \item A timetagger at each of the separated stations Alice and Bob, and a mechanism to synchronize the timetaggers so that they share the same time-base.
    \item A pair of hardware random number generators (RNGs), and accompanying electronic circuitry at each station that provides a reference signal to the timetaggers every time the hardware RNGs emit a usable random bit (this happens once per trial). 
\end{itemize}

\begin{figure}
\includegraphics[width=\linewidth]{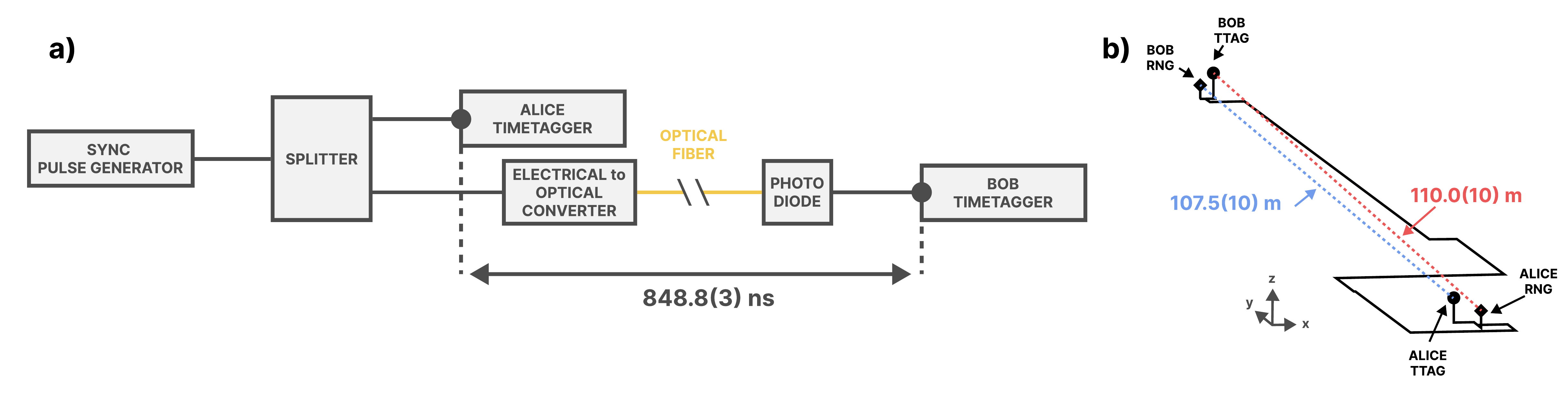}%
\caption{a) Schematic of the synchronization electronics used to establish a common time-base between the separated timetaggers at the Alice and Bob station. b) Schematic of the distance measurements between the Alice (Bob) RNG and the Bob (Alice) timetagger (in black), along with the inferred free-space separation (in red and blue). \label{trusted measurements}}
\end{figure}

The spacelike separation we enforce is that the outcome of a trial be recorded to the timetagger at the Alice (Bob) station before any information about the random bit choice at the Bob (Alice) hardware RNG (traveling at the speed of light in vacuum) could reach the timetagger at the Alice (Bob) station. We assume that once an electrical signal enters the timetagger, no external signal can change it, and that it is subsequently recorded truthfully.
To enforce this condition, we first establish a common time-base for the separated timetaggers. We do this by employing an electrical synchronization circuit that produces a voltage pulse about once every \qty{10}{\second}. This pulse is split and recorded on both timetaggers. By measuring the latency from the splitter to the Alice and Bob timetaggers, we can establish a common time zero between the separated timetaggers. The details of circuit and the relevant measurement are presented in Fig~\ref{trusted measurements}. The uncertainty of the measurement is estimated by a propagation of standard uncertainties of the individual measurements that make up the final measurement in the figure. We also performed a second consistency check on this time delay, and found that these two measurements agreed within \qty{0.26(64)}{ns}, which is within zero to within the estimated uncertainty. 

We also need to characterize and trust the constancy of the shortest physical distance from the Alice RNG to the Bob timetagger and from the Bob RNG to the Alice timetagger. These are the distances that hypothetical signals carrying information about the settings choices from one station to the other would need to travel through. We rely on measuring a set of orthogonal spans using a tape measure to establish these distances. The orthogonal spans are indicated by the black lines in Fig.~\ref{trusted measurements}. The sources of uncertainty in this measurement are two-fold. There is uncertainty in each of the individual measurements themselves, and also uncertainty in the orthogonality of each span with respect to the others. Efforts were made to ensure that the spans were parallel to the building during each measurement, but the possibility of the building not being exactly “square” remains. To account for these uncertainties, we run a Monte-carlo simulation with 1 million individual estimates. For each estimate, we sample the measurements of each individual span from a normal distribution with a standard deviation of \qty{1.55}{mm}, which we determine to be the measurement uncertainty on our tape measure. Simultaneously, we allow the solid angles between the individual spans to vary with a standard deviation of \ang{4.5}. The results of such a simulation follow a distribution that is approximately normal, and the combined standard uncertainty for each of these measurements is \qty{1.0}{m} (depicted in Fig.~\ref{trusted measurements}). Note that the best estimate of the distance actually depends weakly on estimated the angular uncertainty, and the probability distribution of angles. For our uncertainty estimation, we employed a projected normal distribution to model the angular uncertainty.

Finally, for every trial performed, we must establish the start of the RNG, or the first moment when any information about the random bit choice of the RNG could be revealed to outside (environment). We do this by characterizing the hardware RNGs and the electronic circuits that produce a trial marker. The trial marker is an electric pulse that is recorded on the timetagger which signals the start of the random number choice. The latency of the start of the RNG with respect to the trial marker is 31.0(8) ns at Alice and 24.6(3) ns at Bob. We employ two hardware RNGs at Alice, and while we only need to consider the RNG that starts earlier, we also need to trust that the second RNG does not start much earlier than our measurements indicate.

\subsection{Measurements on untrusted devices}
\begin{figure}
\includegraphics[width=\linewidth]{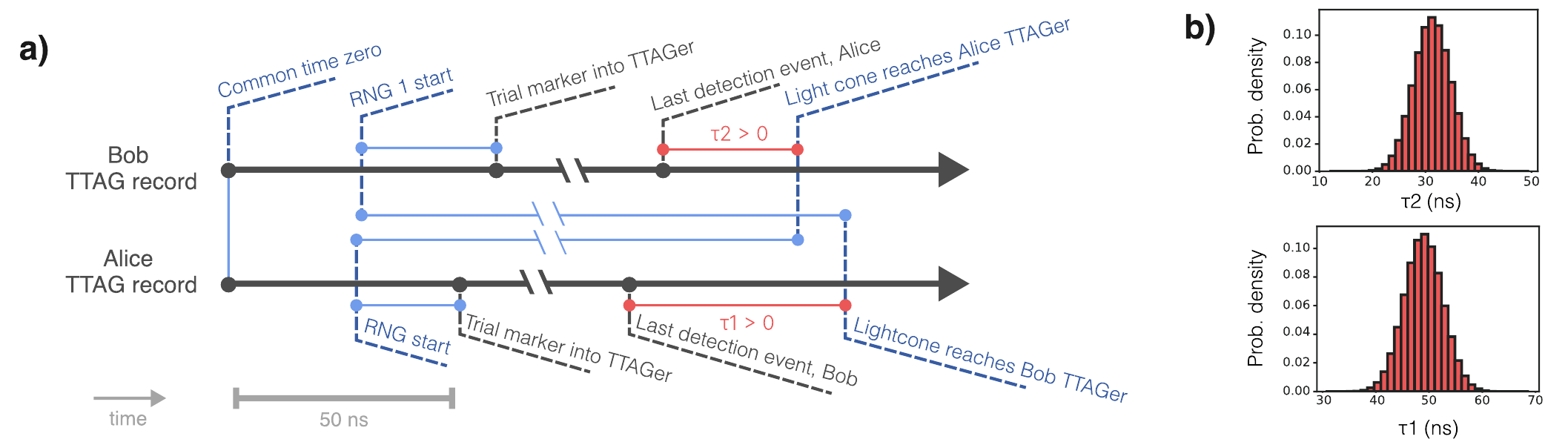}%
\caption{a) Timing diagram representing the timings from the measurements on untrusted devices. The solid blue lines represent measurements we trust remain constant throughout the course of the experiment. The important events on both timetagger records are denoted by black circles. The requirements to be satisfied for space-like separation are that the last detection event at Alice (Bob) happens before the lightcone from the Bob (Alice) RNG reaches the Alice (Bob) timetagger. These requirements are captured by $\tau1 >0$ and $\tau2 > 0$, represented by the red lines in the diagram. These time differences can be estimated based on events from the timetagger records and the trusted measurements (blue lines). b) A Monte Carlo simulation representing the expected spread in the $\tau1$ and $\tau2$ values in nanoseconds, based on the 7434 worst-case $\tau1$ and $\tau2$ from each successful pulse and the combined uncertainty from all timing and distance measurements. The fact that both these measurements are significantly greater than zero after taking uncertainties into account means all reported trials were performed with significant space-like separation. \label{untrusted measuements}}
\end{figure}

The untrusted devices include much of the electronics and photonics to achieve a Bell violation. For establishing spacelike separation, it suffices to ensure that the last detection event---which corresponds to the last outcome from the untrusted devices---is recorded to our trusted Alice (Bob) timetagger before any light-speed signal from the Bob (Alice) RNG could reach and potentially modify the signal. We can establish the start of this hypothetical light-speed signal from each RNG based on when the trial marker is recorded on each timetagger, and a trusted measurement of the start of the RNG with respect to the trial marker (discussed in the previous section), as indicated in Fig.~\ref{untrusted measuements}. Because the last detection event is also recorded on the timetaggers, we can then estimate the times (t1 and t2) from the last detection event to when any information about the settings choices could reach the timetaggers. These estimates use the measurements of the distances discussed in the previous section, and as indicated in Fig.~\ref{untrusted measuements}. In order to demonstrate that all performed trials were spacelike separated, it suffices to show that the t1/t2 estimates for the worst case trials are significantly greater than zero. Fig.~\ref{untrusted measuements} shows a Monte-Carlo simulation for all worst case t1 and t2 times from October 27 2023 to December 6 2023, based on the uncertainty of our best estimate. The results for these days are: $\mathrm{t2} = \qty{31.3(3.5)}{\nano \second}$ and $\mathrm{t1} = \qty{49.0(3.6)}{\nano \second}$, showing that the experiment operated with significant spacelike separation.

\section{Entangled photon source and measurement stations}

The NIST experiment uses an entangled pair source largely similar to prior work~\cite{shalm_strong_2015, shalm_device-independent_2021-1} to probabilistically generate a polarization entangled state close to $0.383
|\mathrm{HH}\rangle + 0.924
|\mathrm{VV}\rangle$ where the H and V represent horizontal and vertical polarizations. 

The state is generated via spontaneous parametric downconversion in one of two paths through a periodically poled potassium titanyl phosphate (PPKTP) crystal placed inside a Mach-Zehnder interferometer. To characterize the quality of our source, we attempt to prepare the singlet state---which is ideally perfectly anti-correlated in all bases---and measure the anti-correlation visibility in 360 equally spaced bases around the Bloch sphere. The visibilities we measured were between $0.9991(1)$ and $0.9907(3)$, with the highest visibilities measured close to the horizontal/vertical polarization basis, and decreasing as we move towards the diagonal/anti-diagonal and left/right-circular bases. 
For the experiment, the optimal entangled state ($ 0.383|HH\rangle + 0.924|VV\rangle$) and measurements are found via a numerical maximization of violation of the Clauser-Horne (CH) inequality~\cite{shalm_strong_2015} that takes losses, background counts and imperfect state visibilities into account. 

As shown in Fig. 2 of the main paper, the entangled source is pumped with $\approx \qty{6}{nJ}$, \qty{25}{\nano \second} pulses from a gain-switched laser operating at \qty{80.00(1)}{\mega \hertz} and centered at a wavelength of \qty{774.3(2)}{\nano \meter}. This allows us to produce close to $2.2\times 10^5$ entangled photon pairs per second at $\approx \qty{1550}{nm}$. The reliability of the turnkey laser over a Ti:sapphire laser allows us to operate the experiment for long periods without manual intervention.  

The generated photons are then coupled into SMF-28 single-mode fibers with thermally expanded cores designed for more efficient free-space fiber coupling and sent to the remote stations, Alice and Bob. Two pairs of mirrors in motorized mounts before each fiber coupling stage are used to automatically optimize alignment when necessary. The highest symmetric efficiency we have observed with this setup is $0.8283(6)$, but the efficiency is only around $\approx 0.81$ during the automated operation described in this work. Waveplates in motorized mounts also automatically pre-compensate the polarization drifts introduced by the fibers. At these remote stations, the entangled photons are measured in one of two bases---Alice chooses between $a = \ang{6.7}$ and $a^{\prime} = \ang{-29.26}$ and Bob chooses between $b = \ang{-6.7}$ and $b^{\prime} = \ang{29.26}$, where the angles are relative to a horizontally oriented polarizer.

These choices are made independently and at random via hardware random number generators at each station producing random bits based on phase diffusion in a laser diode~\cite{abellan_generation_2015}. The choices at the Alice station are additionally XORed with random bits from a photon sampling random number generator~\cite{wayne_post-processing-free_2018} to minimize the possibility of common correlations in the input randomness. The random bits are then fed to Pockels cells that rapidly switch the projective measurements performed on the photon between the two polarization bases at each station.  The photon is then coupled into a single-mode fiber that terminates at a superconducting nanowire single photon detector (SNSPD) with high system detection efficiencies~\cite{reddy_superconducting_2020}. A detection or no detection event at the SNSPD constitutes a projective polarization measurement of the photon in the basis decided by the Pockels cell and a set of three waveplates, as in Fig.~2 of the main paper.

The sync signal---consisting of one electrical pulse for every $320$ pump periods---distributed from the source acts as a master clock for the experiment. The Pockels cells are triggered on the sync signal, at a rate of $2.5\times10^5\ \mathrm{Hz}$ ($\qty{80}{MHz}/320$), and stay on for around $\qty{200}{ns}$ or $14$ pump pulses. Each triggering of the Pockels cells defines one Bell trial. Each Bell trial takes in two bits to make the random settings choices at Alice and Bob, and puts out one bit at each measurement station encoding the detection or no-detection of a photon. In our experiment, only about $4\%$ of the trials coincide with an entangled pair emitted from our probabilistic source, the rest are empty and will deterministically result in two no-detection events (other than a small probability of background count detections). This number is achieved thanks to the aggregation of 14 pulses that allows us to boost the probability of an entangled pair in a trial from the probability of an entangled pair from a single pump pulse ($P_{pulse} = \frac{1}{363}$).

\section{Long term performance of device-independent RNG}
\label{sec:longterm}
\subsection{Adaptation of theoretical protocol}

The experiment in this work was designed to run for long periods of time without manual intervention, and required an automated updating of the probability estimation factors (PEFs) used to certify entropy in the outputs, to account for any long term drifts in the hardware. 

To explain the PEF formalism, it is useful to define a few terms. Let $Z=XY$ be a random variable denoting the per-trial inputs $X, Y$ to Alice and Bob, and $C=AB$ be a random variable denoting the outputs $A, B$ from Alice  and Bob. Also let the lowercase $c$, $z$, etc. denote the possible values that the random variables $C$, $Z$, etc. take on.
Then, as defined in Y. Zhang et al. (2018)~\cite{zhang_certifying_2018}, a PEF with power $\beta > 0$ is a function $F: cz \mapsto F(cz) \geq 0$ such that $\sum_{cz} F(cz)\sigma(cz)\sigma(c|z)^\beta \leq 1\  \forall\  \sigma \in \mathcal{T}_{CZ}$. Here $\mathcal{T}_{CZ}$ is a trial model consisting of all allowable (per-trial) probability distributions of $CZ$. The trial model we employ in this work is similar to the one employed in prior work~\cite{zhang_experimental_2020}, and detailed in section VIII of the the arXiv version of E. Knill et al. (2017)~\cite{knill_quantum_2020}. It satisfies no-signaling constraints, Tsirelson's bounds and allows for an adversarial bias $\epsilon_b$ in the settings where---in our work---$\epsilon_b \leq 10^{-3}$.

Given a set of trial-wise PEFs $F_i$, we can define $T_0 = 1$ and $T_i = \prod_{j=1}^{i} F_j(C_jZ_j)$. The accumulated product $T_n$ after $n$ trials can be related---conditional on it being larger than a predetermined quantity---to a lower bound on the $\epsilon_h$-smooth conditional min-entropy $H_{min}^{\epsilon_h}(\mathbf{C}|\mathbf{Z}E)$ with respect to classical side information $E$, as shown in E. Knill et al. (2017)~\cite{knill_quantum_2020}. Here, the bold letters denote a sequence of inputs ($\mathbf{C}$) and outputs ($\mathbf{Z}$) from a sequence of trials.

In practice, we found that it was sufficient to update the PEFs about once a day during the course of the demonstration. At the start of each day and after a fridge recycle, 20 minutes worth of data from the previous day was used as calibration to construct new PEFs. The procedure used to compute the new PEFs was largely similar to the one detailed in a prior work~\cite{zhang_experimental_2020}, 
 and is summarized below.

The calibration data is first used to estimate the per-trial input-conditional distribution $\nu(C|Z)$.  The distribution is estimated under the independent and identically distributed (i.i.d.) assumption subject to no-signaling and Tsirelson's bound constraints, by maximum likelihood. This ensures we find the most likely input-conditional distribution consistent with the experimental model of a quantum system, in the event that finite statistics effects result in a violation of these constraints. This input-conditional distribution serves as the reference distribution for subsequent optimization of the PEFs. In particular, the distribution $\nu(C|Z)$ is the unique solution to the convex optimization problem

\begin{equation}
\begin{aligned}
      \max_{\mu(C|Z)} &\sum_{cz} n_{cz} \log(\mu(c|z)) \\
    \operatornamewithlimits{with}\  & \mu(C|Z) \in \mathcal{T}_{C|Z}.
\end{aligned}
\end{equation}
Here, $n_{cz}$ is the number of calibration trials with $C=c$ and $Z=z$, and $\mathcal{T}_{C|Z}$ is the convex polytope of conditional probabilities satisfying Tsirelson's bounds and no-signaling constraints~\cite{knill_quantum_2020}. This convex optimization is solved via ECOS~\cite{domahidi_ecos_2013} through a Python implementation~\cite{noauthor_buff_nodate}. 
 
Next, a PEF can be obtained by optimizing on this calibration data. While a fixed PEF is always valid, using PEFs optimized for more recent calibration data results in better performance. To obtain the new PEF, we maximize a quantity that is related to negative logarithm of the probability of the most likely bitstring that could be obtained from any of the probability distributions in the experimental model $\mathcal{H}(\mathcal{T}_{CZ})$, which can be obtained by chaining the trial model $\mathcal{T}_{CZ}$~\cite{zhang_certifying_2018} (see section I in the supplementary information of Y. Zhang et al. (2020)~\cite{zhang_experimental_2020} for a discussion on the construction of trial models). In other words, we attempt to maximize a lower bound on $H_{min}^{\epsilon_h}(\mathbf{C}|\mathbf{Z}E)$. In particular, the first optimization step, given a power $\beta$, is 

\begin{equation}
    \begin{aligned}
      \max_{F(CZ)} &\mathbb{E}_{\nu}\big(\log_2(F(CZ))\big) \\
    \operatornamewithlimits{with}\  & \sum_{cz} \mu(cz)F(cz)\mu(c|z)^\beta \leq 1\:  \forall\:  \mu(CZ) \in \mathcal{T}_{CZ}, \\
    &\:\:\:\: F(cz) \geq 0 \: \forall \: cz.
\end{aligned}
\end{equation}

Finally, a numerical optimization over $\beta$ is performed, with the expected number of trials needed to certify the requested bits as the objective to be minimized.

\begin{figure}
\includegraphics[width=\linewidth]{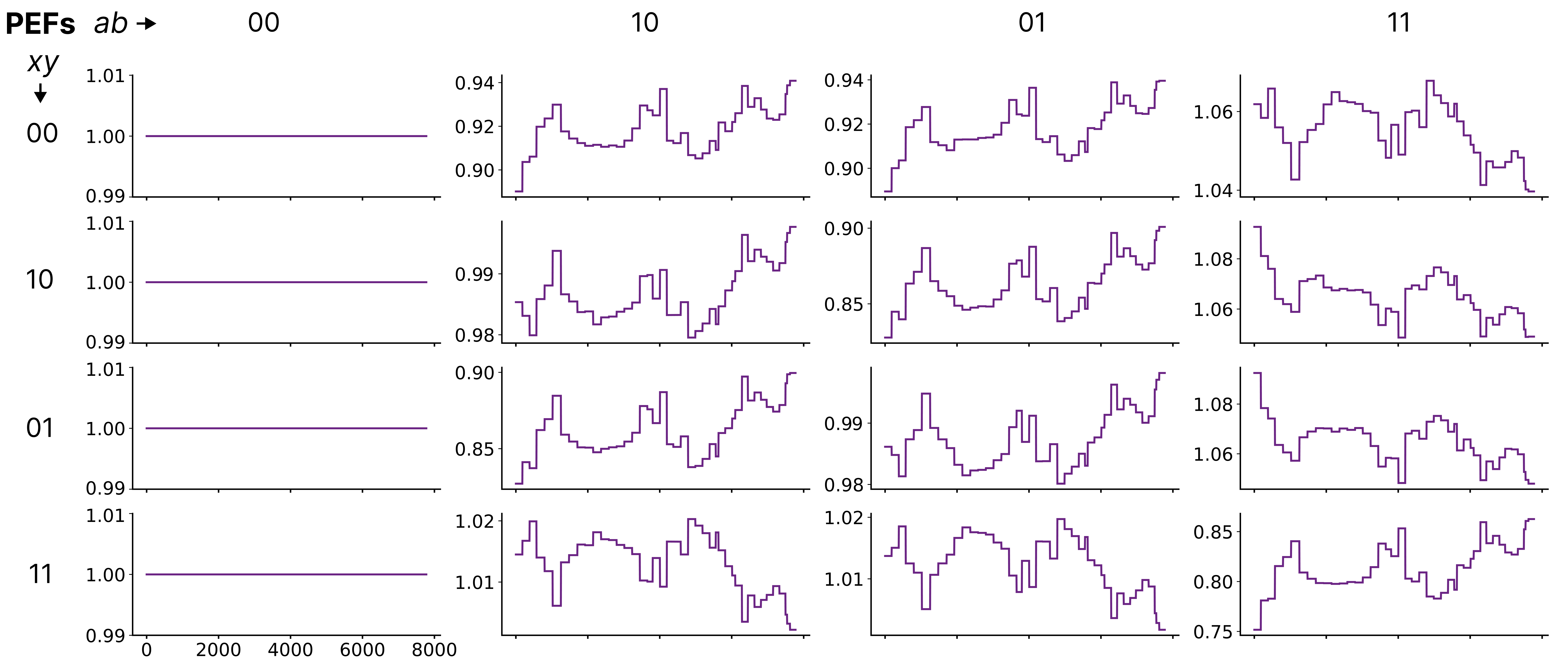}%
\caption{The 16 elements of all the PEFs committed to during the course of the experiment, all plotted as function of the pulse index, which is the number assigned to pulses from the CURBy-Q chain. They are computed from calibration data. $a,b$ are the outputs and $x,y$ are the inputs from the Alice, Bob stations during the Bell experiment.\label{pef figure}}
\end{figure}

\begin{figure}
\includegraphics[width=\linewidth]{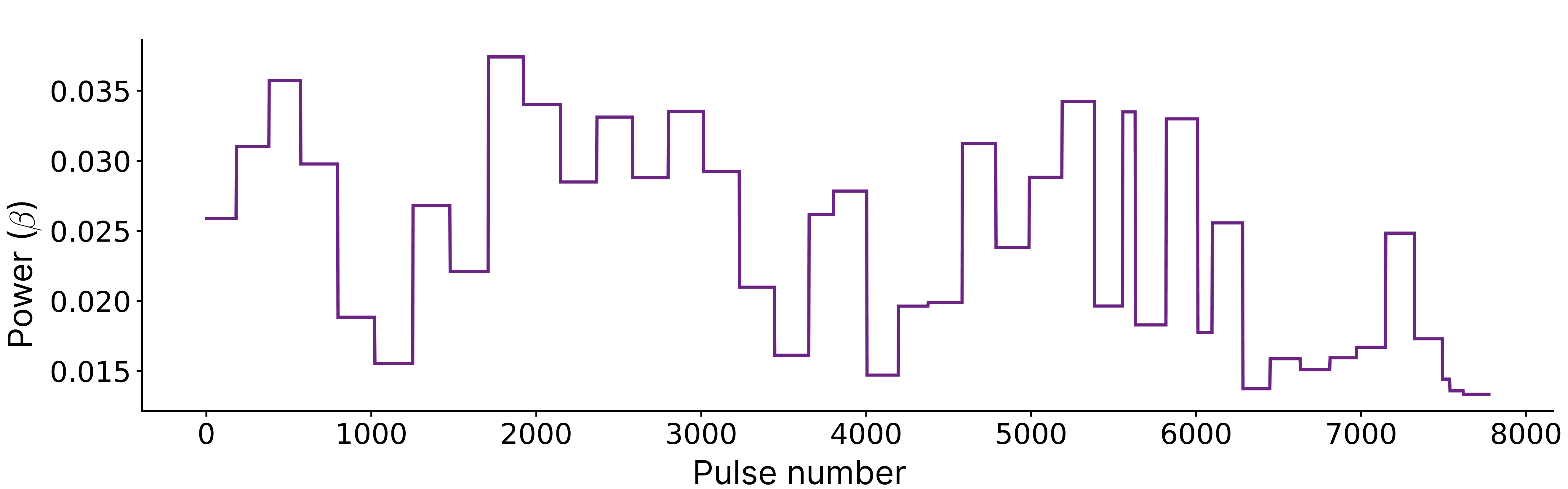}%
\caption{Plot of the PEF power $\beta$ committed to before every request for fresh data from the DIRNG experiment. The pulse number is the ordinal number of the corresponding random pulse request on the CURBy-Q hash chain.\label{beta figure}}
\end{figure}

The results of the PEFs and powers ($\beta$) used over the course of the demonstration are plotted in Fig.~\ref{pef figure} and ~\ref{beta figure} respectively.

\subsection{Experimental performance}

The experiment in this demonstration was run over multiple days, starting at 17:46 October 26, 2023 (UTC) and continued to run intermittently in the following months. For the purposes of this paper, we choose to analyze the experiment starting from 00:03 October 27, 2023 (UTC) to 23:04 December 05, 2023 (UTC). This corresponds to 40 full days from the start of the experiment.

\begin{figure}
\includegraphics[width=\linewidth]{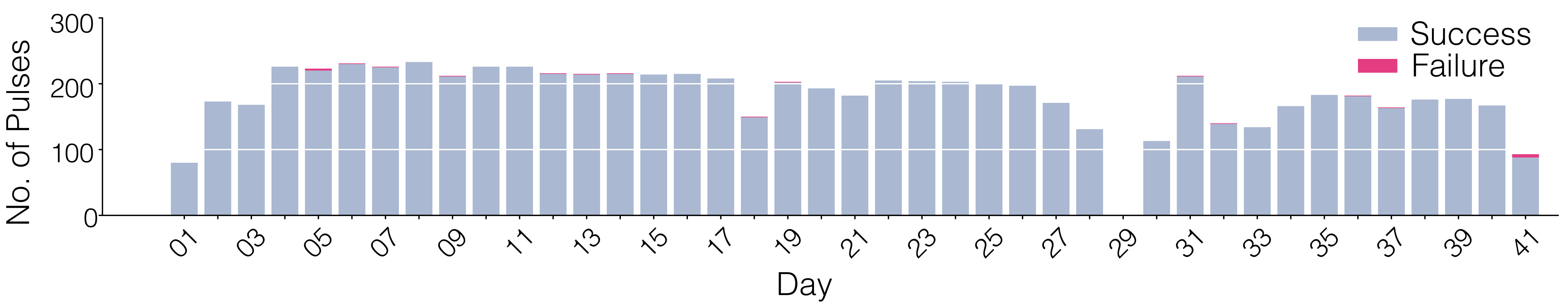}%
\caption{Histogram of the number of random number pulses published each day starting from October 26 2023 (MDT) during the course of the demonstration. The x-axis corresponds to days in local time (MDT). While the analyzed data corresponds to 40 days in UTC, the data is split up into 41 ``local time" day bins.\label{fig:day stats}}
\end{figure}

During this period, the experiment ran continuously except for day 29 (23 November 2023), as indicated in Fig.~\ref{fig:day stats}, when a hardware failure of a computer controlling the cryogenic fridges occurred at the same time as a campus closure at the National Institute of Standards and Technology (NIST).

\begin{figure}
\includegraphics[width=\linewidth]{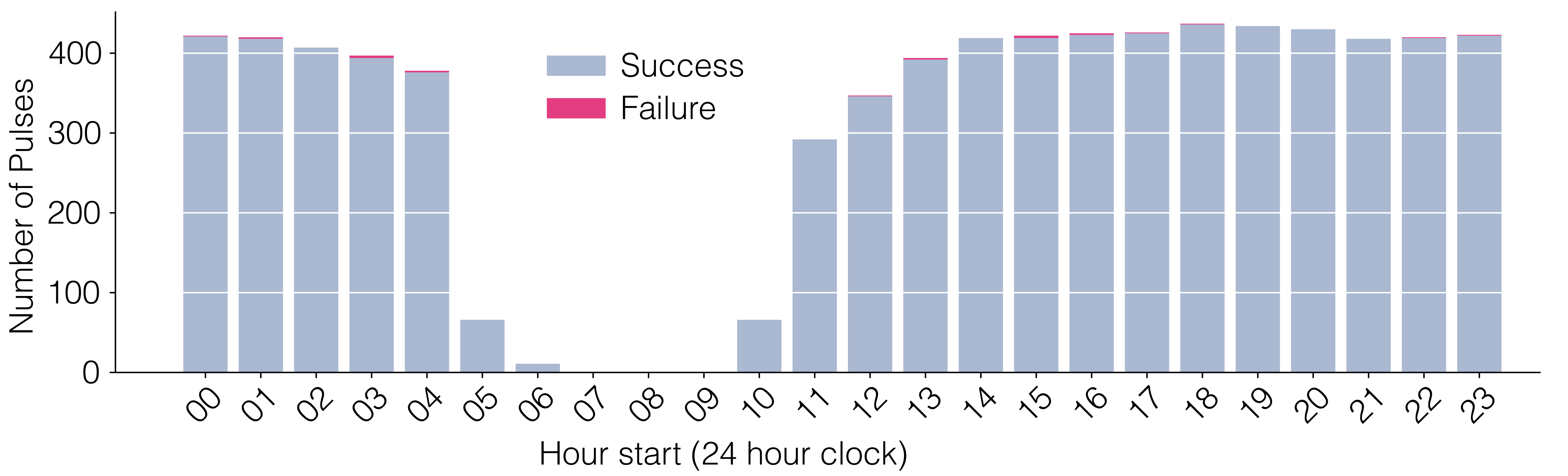}%
\caption{Cumulative histogram of the number of random number pulses served during the course of the hour of the day (indicated on the x-axis). For example, the bar plotted at `00' on the x-axis corresponds to all the pulses served from 00:00 to 00:59 on the 41 (local) days starting from 26 October 2023 (MDT) under consideration. All times are in local experiment time (MDT).\label{fig:hour stats}}
\end{figure}

Over the course of a day, the experiment did not have a 24 hour uptime, as indicated in the main text, and did not supply data when the \textsuperscript{4}He sorption-pumps in the cryogenic fridges housing the superconducting detectors were recycling. This shows up as an interruption of service from about 5 AM to 11 AM Boulder time (MST/MDT) during which no requests are served. This is visible in Fig.~\ref{fig:hour stats}, which is a cumulative histogram of the times when data was returned from the NIST experiment over the 40 days of experimental time.

\begin{figure}[H]
\includegraphics[width=\linewidth]{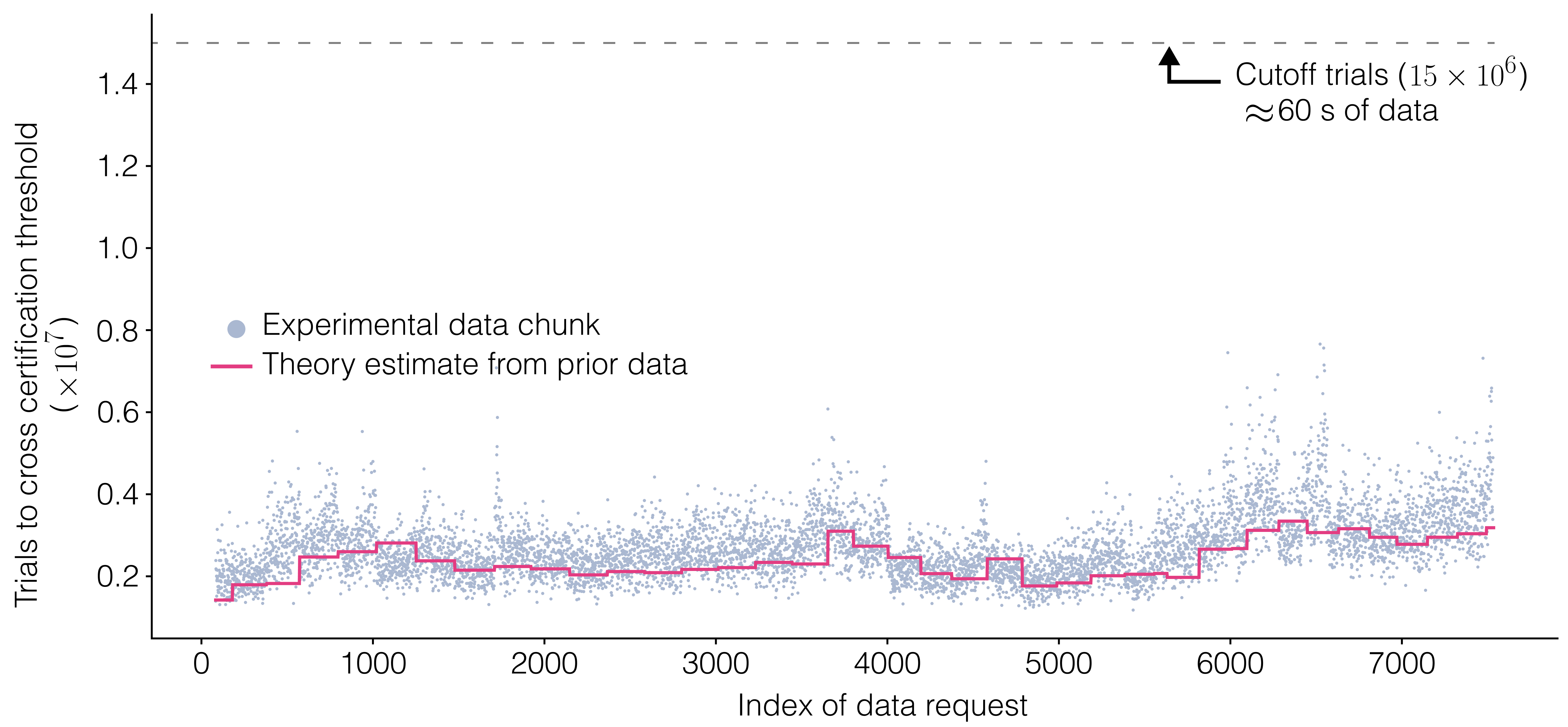}%
\caption{Scatter plot of the minimum truncated trial length at which the running entropy estimate for each of 7434 successful raw data responses crosses the certification threshold (820 bits). Also plotted (solid pink line) is the expected number of trials from the calibration data and the precommitted PEFs (calibration data is from previous day). This is calculated using the formula for $n_{exp}$ in Eq.~\ref{eq: n exp}. The experiment closely follows theoretical expectations from calibration, which is then adapted on a daily basis from experimental data. This arrangement allows us to run the experiment over an extended period of time. The data plotted here corresponds to 40 days starting on 27 October 2023 (UTC), as in figure 4 of the main paper. All of the ``crossing trials" during this period are well below the pre-determined cutoff trial threshold (15 million, grey dashed line in figure). This enables us to achieve a high success rate (99.7\%), limited by hardware glitches that result in invalid data. The index of the pulse corresponds the indexing employed by the CURBy-Q hash chain. \label{fig:all stats}}
\end{figure}

Based on 20 min of calibration data from the previous day, we can estimate the expected number of trials $n_{\mathrm{exp}}$ required to satisfy the standard request for 512 bits of entropy (see section~\ref{sec:randcert}). The $n_{\mathrm{exp}}$ is given by (based on results from E. Knill et al. (2017)~\cite{knill_quantum_2020}),

\begin{equation}\label{eq: n exp}
    n_{\mathrm{exp}} = \frac{\beta\times \sigma_h - \log_2(\epsilon_h)}{\mathbb{E}_{\nu}\big( \log_2(F(CZ))\big) },
\end{equation}

where $\sigma_h$ is the entropy threshold and $\mathbb{E}_{\nu}$ is the expectation according to the distribution $\nu(CZ)$, representing a reference distribution for the experiment from the calibration data (see section~\ref{sec:randcert} for more details on the various parameters).  Fig.~\ref{fig:all stats} is a plot of the expected number of trials as a function of the pulse number, starting from pulse zero on October 26 2023 and continuing on to pulse 7774 published on December 7 2023. Also plotted are the number of trials $n_{\mathrm{cross}}$ when the running entropy estimate $-\log_2(T_n\epsilon_h)/\beta$ (see section~\ref{sec:randcert}) crosses the threshold entropy required for producing 512 bits entropy at the output of the extractor. The pre-determined cutoff trials for the experiment are 15 million, and all the served data over these 40 days is well below that limit. The theoretical estimate closely tracks experimental drift; this is enabled by the adaptive nature of the probability estimation framework used to certify entropy in our experiment. The spikes in the experimental data indicate infrequent drops in experimental performance that increase $n_{\mathrm{cross}}$ with respect to the expected value from the calibration data set. However, all plotted points are still well below the cutoff criterion of 15 million trials (corresponding to \qty{60}{\second} of data) and the experiment performs well enough to meet the demands of the random number beacon throughout the 40 days of this demonstration.

\subsection{Experimental latencies}
For the analyzed data, it took 406 s on average (median latency \qty{267}{\second}) to complete a run of our distributed protocol to generate, extract, and publish 512 random bits. From the time a request for randomness is registered on the CURBY–Q chain, it took NIST \qty{37}{\second} on average (median time \qty{36}{\second}) to register the request. After this, it took a variable amount of time before the experiment was ready to begin collecting the Bell trail data for \qty{60}{\second} and process it. The total time from registration of a request at NIST to completion of local processing was an average of \qty{263}{\second} (median \qty{121}{\second}). The machine running the CURBY-Q chain took an average of \qty{46}{\second} (median time \qty{45}{\second}) to retrieve the data, perform PEF certification, and precommit to a result pulse containing randomness. Because we wait for a new DRAND pulse as input to compute seed bits for the Trevisan extractor, the mean latency before the bits are finally published from this point on was \qty{60}{\second} (median \qty{52}{\second}). Of all of these times, the variance in the time to start collecting Bell data after a request is registered at NIST is the highest: up to \qty{11}{hours} sometimes because the cryogenic fridges housing the superconducting detectors must recharge daily. Upgrading to continuously operational cryogenic fridges would largely mitigate this. 

\section{Randomness Certification and Extraction} 
\label{sec:randcert}
After collection of \qty{60}{\second} of loophole-free Bell test data from the NIST experiment, the data is terminated to $15$ million trials (our pre-determined stopping criterion), and certification of $\epsilon_h$-smooth min-entropy in the outputs is attempted under the PEF framework ($\epsilon_h = 0.8\times 2^{-64}$). If successful, we employ a ``classical-proof" extractor~\cite{mauerer_modular_2012} to extract a uniform bitstring with extractor error $\epsilon_x = 0.2\times 2^{-64}$. Below we present results for soundness of this protocol, based on the proof of Theorem 2 that can be found in the supplementary information of reference~\cite{shalm_device-independent_2021-1}.
 
As mentioned in section~\ref{sec:longterm}, the accumulated product of PEFs $T_n$ is related to a lower bound on the smooth conditional min-entropy $H_{min}^{\epsilon_h}(\mathbf{C}|\mathbf{Z}E)$. To state the exact bound, let $\frac{1}{\mathrm{Rng}(\mathbf{C})} \leq p \leq 1$, and $\{\phi\}$ be the event that $T_n > \frac{1}{p^{\beta}\epsilon_h} $. Define $\kappa = P_{\mu}(\phi)$, where $P_{\mu}$ is the probability according to a probability distribution $\mu$ that is in the experimental model $\mathcal{H}(\mathcal{T}_{CZ})$. Then, as proved in Y. Zhang et al. (2018)~\cite{zhang_certifying_2018}, the $\epsilon_h$-smooth min-entropy of the outputs conditioned on $\phi$ and $\mathbf{Z}E$,
\begin{equation}
    H_{min}^{\epsilon_h}(\mathbf{C}|\mathbf{Z}E; \phi)  \geq -\log_{2}\Big(\frac{p}{\kappa^{1 + 1/\beta}}\Big),
\end{equation}
where the bold letters denote the full sequence of inputs (\textbf{C}) and outputs (\textbf{Z}) from 15 million trials.

In our demonstration, the event $\phi$ is interpreted as successful certification of min-entropy during a protocol run. Upon a success, we use this conditional entropy bound with the TMPS extractor to produce the output bits that are subsequently published.

The protocol we implement in this demonstration is detailed in Algorithm~\ref{alg:randgen} below. 

\begin{algorithm}[H]
\caption{Beacon randomness generation protocol}\label{alg:randgen}
\begin{algorithmic}[1]
\Procedure{random number generation}{$\epsilon_x, \epsilon_h, \sigma$}\Comment{$\epsilon_x + \epsilon_h = \epsilon = 2^{-64}, \ \sigma = 512 $}
%\Ensure $y = x^n$
\State $F \gets$ PEF from calibration data  \Comment{PEF precommitment}
\State $n \gets 15 \times 10^6$ \Comment{Stopping criterion (number of trials)}
\State $\sigma_h \gets \lceil \sigma + 4\log_2(\sigma) + 6 -4\log_2(\epsilon_x)\rceil $ \Comment{Entropy threshold for Bell test outputs (bits)}
\Function{find prime}{$m, k, \epsilon$}
%\Require $p \in \{P}$
\State \multilinestate{$w \gets$ smallest $p$ 

such that $p \in \mathbb{P}$,\  $p > 2\lceil \log_2(4mk^2/\epsilon^2)\rceil$ 

where $\mathbb{P}$ is the set of prime numbers}
\RETURN{} $w$
\EndFunction
\State $w \gets$ \Call{find prime}{$2n, \sigma, \epsilon_x$}
\State $l \gets w^2\times\max \Big( 2 , 1 + \Big\lceil\frac{\log_2(\sigma - e) - \log_2(w - e)}{\log_2(e) - \log_2(e -1)}\Big\rceil \Big)$ \Comment{Length of independent seed}
\State $\mathbf{c},\mathbf{z} \gets$ \Call{Run NIST DI-RNG}{n} \Comment{Get \qty{60}{s} of trial data from NIST Loophole-free Bell test}
\State$T_n \gets$ $\Pi_{j=1}^n F(c_j,z_j)$ \Comment{Compute the accumulated PEF product}
\If {$(T_n\epsilon_h)^{-1/\beta} > 2^{-\sigma_h}$} \Comment{Protocol failed}
\State $\mathcal{Q}_P \gets 0$
\State $\mathcal{Q}_X \gets \varnothing$
\State $\mathcal{Q}_S \gets \varnothing$
\RETURN{} $\mathcal{Q}_P,\mathcal{Q}_X, \mathcal{Q}_S  $
\Else
\State $\mathcal{Q}_P \gets 1$
\State $s_{\leq 512} \gets$ \Call{DRAND}{next} \Comment{Obtain next 512 bit pulse from drand.love beacon}
\State $s_{\leq l} \gets$ \Call{SHAKE256}{$s_{\leq 512}$} \Comment{Expand to $l$ seed bits with the SHAKE256 algorithm}
\State $\mathcal{Q}_X \gets \mathcal{E}(\mathbf{c}, s_{\leq l}, \log_2(T_n\epsilon_h)/\beta, \sigma, \epsilon_x) $ \Comment{TMPS extractor~\cite{mauerer_modular_2012}}
\State $\mathcal{Q}_S \gets s_{\leq l}$
\RETURN{} $\mathcal{Q}_P,\mathcal{Q}_X, \mathcal{Q}_S  $
\EndIf
\EndProcedure
\end{algorithmic}
\end{algorithm}

This protocol is proved to be $(\sigma, \epsilon)$ sound with respect to an external entity $E$ in control of the devices in Theorem 21 of E. Knill et al. (2017)~\cite{knill_quantum_2020}, in the sense that for all $\mu \in \mathcal{H}(\mathcal{T}_{CZ})$, there exists a distribution $\nu_E$ of $E$ such that
\begin{equation}
    \mathrm{TV}\big(\mu[\mathcal{Q}_X \mathcal{Q}_SE| \mathcal{Q}_P = 1], \mathrm{Uniform}_{\mathcal{Q}_X}\otimes\mu[\mathcal{Q}_S]\otimes\nu_E\big)\mathbb{P}(\mathcal{Q}_P = 1) \leq \epsilon,
\end{equation}
where $\mathrm{Uniform}_{\mathcal{Q}_X}$ is the uniform distribution over $\mathcal{Q}_X$, and $|\mathcal{Q}_X| = \sigma$. 
The total-variation distance between distributions $\mathrm{TV}(\mu, \mu')$ is the largest difference in probabilities assigned to the same event by $\mu$ and $\mu'$, and is given by

\begin{equation}
    \mathrm{TV}(\mu, \mu') = \frac{1}{2} \sum_x |\mu(x) - \mu'(x)|.
\end{equation}

\section{The Twine Protocol}

\subsection{Overview}

The data structure underpinning CURBy records is created using the Twine protocol developed in tandem, with CURBy as its first use-case. The Twine protocol specifies a method by which a decentralized group of independent parties can cooperate to produce a ledger of immutable ordered data, and it does this without requiring any kind of consensus mechanism. The authorship (i.e. provenance) of the data is verifiable using digital signatures and public key cryptography. Every chunk of data is linked via hash-chaining and the resulting data structure, which we call a Tapestry, forms a directed acyclic (hyper)graph (DAG). As a consequence, the data assumes a partial ordering in a cryptographically verifiable way, while also ensuring its integrity and provenance. The data becomes immutable and non-repudiable by any party---even its creator---due to the intertwining of data produced by independent sources. We detail the Twine protocol in the following sections.

\subsection{Cryptographically Secure Time Ordering}

Timestamps, like those described by the ISO 8601 standard (eg: \code{”2024-10-31T13:59:59Z”}), are the most common way to record time digitally. These are uncertified declarations of time, and their authenticity cannot be ensured. A common way of adding credibility to a timestamp associated with some data is to get the timestamp certified for that data by a trusted authority, as is done within the RFC 3161 Standard~\cite{zuccherato_internet_2001}. At its core, this involves sending the data one wishes to timestamp to a trusted authority, which in turn returns a digital signature of the data combined with a timestamp. This digital signature serves as proof---reliant on that authority---that the data was created no later than that time.

In practice, instead of sending the raw data, a hash of the data is sent instead. The hash serves as a cryptographic fingerprint of the data that reveals no information about the data itself.

There are some drawbacks to this method of timestamping, however. Firstly, one must trust the integrity of the timestamping authority, and trust that they are neither conspiring to produce inauthentic timestamps or leaking their cryptographic keys to other parties. Secondly, the timestamp only represents an upper bound after which the data could not have been produced. For example, one could imagine someone creating a piece of data and holding it for years before timestamping it in this way.

To prove that a piece of data must have been created after a specific time, one can incorporate the signature (or a hash of the signature) of a prior certified timestamp into the data in question, since this could not have been known prior to signing. In other words, the data becomes a derivative of prior data, making its relative order unambiguous. This is the central idea behind hash-linking, where the hash of some previous data is incorporated into the next data, and so on.

\begin{figure}
    \centering
    \includegraphics[width=0.75\linewidth]{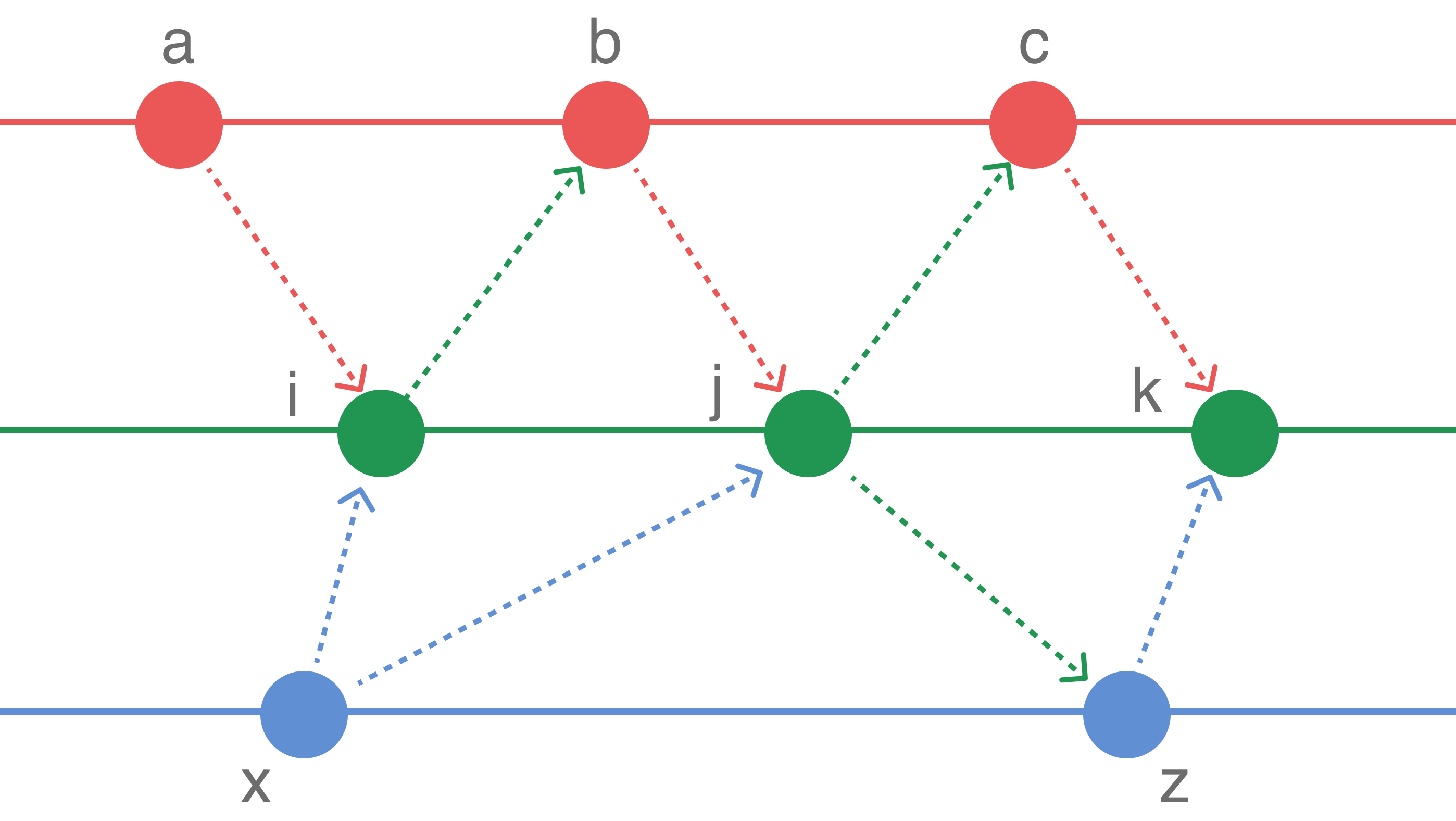}
    \caption{Hypothetical example of a Twine data graph involving three independent parties running the Twine protocol creating pulse data structures shown as circles. Time increases horizontally, left to right. Each pulse is signed by the author, and in addition to including the hash of the previous pulse on their chain, they also include hashes of the previous pulse of their ``neighbors”. Taking the pulses \code{b}, \code{i}, and \code{j} for example, verification of their hash links will prove that pulse \code{b} came after pulse \code{i}, but before pulse \code{j}. Additionally, this extends to indirect hash-linking also, and there is a proof (via \code{i}) that \code{x} came before \code{b}. In general, if a path of hash-links can be found between any two Pulses, an unambiguous order can be proven.}
    \label{fig:toy-model}
\end{figure}

The Twine protocol expands on this foundation, and instead of a client requesting a certified timestamp from an authority, every entity participating in the protocol certifies the timestamps of other participants. In other words, one does not need to place all their trust in a single authority’s timestamp, since that timestamp is in turn certified by another authority, and the ordering of the data is effectively unforgeable due to the properties of hash-chains. See Fig.~\ref{fig:toy-model} for an example of a Twine instance. 

\subsection{Data Structures}

There are two core data structures for Twine records: pulses and chain metadata. Pulses compose the bulk of the Tapestry and contain the information relevant to the use case. In the case of CURBy, for example, the pulses record the information about randomness generation, Bell test execution, and so on. The Pulses are logically grouped as chains, which act as proxies for ownership. Generally, a chain is produced for a single purpose or use case by a single process and owner. Every pulse of every chain is also linked (via hash-linking) to pulses of other chains, thus forming the larger Tapestry. Chain metadata stores meta information about the construction of the chain, including the public key for provenance verification. 

Hash-linking of the data structures is facilitated by the process of content addressing. With content addressing, data is referenced by its hash instead of its location using a special identifier called a Content ID (CID)~\cite{noauthor_ipldipld_nodate}. Content addressing and CIDs were established as part of the distributed peer-to-peer filesystem IPFS, however IPFS is not necessary for Twine data. 

A CID is constructed using the hash of serialized data, meta-information about the serialization method, hash algorithm, format, and version of the CID itself. This self-descriptive CID is used as both an identifier for retrieval of the original data, as well as a checksum to verify its integrity. Since CIDs contain a hash of the data, they can be used for the purposes of hash-linking.

The following is an example of a CID using the SHA3-512 hash function formatted as a base32 string:
\begin{figure}[H]
\centering
\begin{minipage}{0.5\textwidth}
``bafyriqa5k2d3t3r774geicueaed2wc2fosjwqeexfhwbptfgq7rcn5m
wucnhfeuxu2nxbrch3rl6yqjlozhuswo5ln3xwjm35iftt3tpqlcgs"
\end{minipage}
\end{figure}

The pulse and chain metadata structures follow the IPLD data model~\cite{noauthor_ipldipld_nodate} for the purpose of generating CIDs. The IPLD specifications are very open-ended and allow for the use of a variety of hash functions and serialization methods to derive CIDs. The data produced by CURBy uses the SHA3-512 hash function and DAG-CBOR serialization.

Pulse data structures are comprised of several fields which are listed below along with their functional purpose. 

\begin{itemize}
    \item ``chain": A CID of the chain metadata which provides an immutable reference to retrieve the chain metadata (elaborated on later).
    \item  ``index": A numeric index which monotonically increases with each subsequent pulse published as part of its chain. This is an alternate way of identifying a pulse on a specific chain, but it is not appropriate as a secure reference.
    \item ``links": A list of links to pulses on the same chain, which is a list of CIDs. The first link is always to the previous pulse, and the other links serve to facilitate rapid traversal to much earlier pulses.
    \item ``mixins": A list of links to pulses on other chains, which is a list of chain/pulse CID pairs. These create connections external to the current chain, and weave together the Tapestry.
    \item ``payload": Arbitrary data which contains the content relevant to the purpose of the chain, fully specified by the entity constructing the pulses.
    \item ``specification": A specification string, which contains version information about the Twine schema and optionally protocol and versioning information about the pulse payload.
    \item ``signature": The above fields are serialized, hashed, and signed, producing the signature as a JSON Web Signature.
\end{itemize}

All fields (including the signature) are serialized and hashed to produce a CID for that pulse. Chain metadata has a similar composition with some different fields:

\begin{itemize}
    \item ``key": The public key in JSON Web Key format to use for provenance verification.
    \item ``links\_radix": An integer describing how the pulse “links” fields is constructed.
    \item ``meta": Arbitrary data to help describe the use-case of the chain.
    \item ``source": A string identifier representing the owning authority.
    \item ``specification": (same as in the pulse)
    \item ``signature": (same process as described for the pulse)
\end{itemize}

As with the pulse, the above fields are used to create the chain CID.

\section{The CURBy Network}

\begin{figure}
    \centering
    \includegraphics[width=0.75\linewidth]{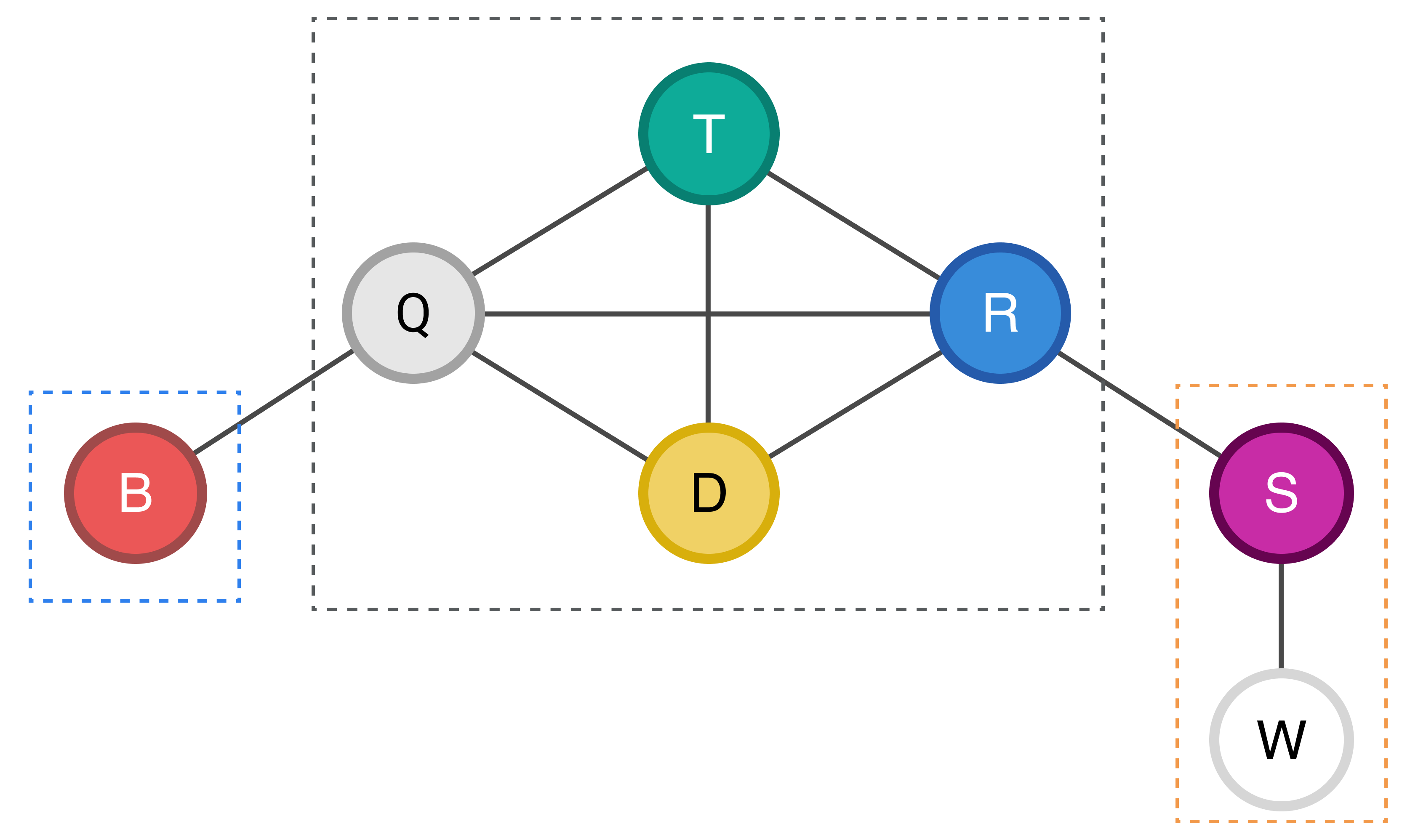}
    \caption{Figure of chain connectivity via hash-chaining. Chains are represented as circles. Identified from left to right, top to bottom as follows: Bell Experiment, CURBy-Q, Time-stamping, DRAND, RNG, Seam, Weather. Black solid lines represent mutual hash inclusion of pulses belonging to those chains. The dotted areas represent geographical locations of the source processes, from left to right: NIST, CU Boulder, Cloudflare workers (the cloud).}
    \label{fig:curby-network}
\end{figure}

The CU Randomness Beacon project is composed of several independently operating and geographically distinct processes. Each process records its data to a unique Twine chain and uses that collective ledger for communication. A visual representation of all chains in the CURBy project is shown as Fig.~\ref{fig:curby-network}.

Each chain has a distinct purpose and protocol, with some depending on the payload of others to advance. There is no direct communication between the processes creating the Twine chains. Instead, they relay information through the central data store, which is accessed using the same HTTP API that is available to the general public for consumption and verification. All chains use the SHA3-512 hash algorithm and all but two use RSA 256 signatures with a 4096 bit modulus. The Seam Chain and the Weather Chain instead use ES256 (P-256) signatures.

\subsection{CURBy-Q and Bell Experiment Chains}

The two most closely dependent chains are the Bell Experiment and CURBy-Q chains. The CURBy-Q chain is tasked with requesting bell experiment data and then performing the certification and extraction process detailed in the section~\ref{sec:randcert} above to produce 512 certified uniform random bits. 

\subsection{DRAND Chain}

The DRAND chain sources its payload from the DRAND randomness beacon pulse data. Its purpose is to act as an independent source of randomness for use as the seed for the Trevisan extractor. The process of obtaining the latest DRAND data and inserting it into a Twine chain is currently run by CU computers, but ideally this would be done by an independent party to further decrease the possibility of malicious tampering. 

\subsection{Time-stamping Chain}

The Time-stamping chain is tasked with providing an independent corroboration of the timing of pulses. The process hashes all mixin data (external pulse CIDs) and uses the \href{http://freetsa.org}{\url{freetsa.org}} time-stamping service to obtain a certified timestamp following the RFC 3161 TSA protocol. The relevant data to verify this certification is recorded in the payload. This provides proof of the timing of all mixins included in the pulse.

\subsection{PRNG (NIST Beacon Inspired) Chain}

The PRNG chain produces regular pulses of 512 pseudorandom bits every 60s. The protocol is adapted from the NIST Randomness Beacon protocol~\cite{kelsey_randomness_2019}, with some modifications. The output randomness meant for public use is the hash digest portion of the pulse’s CID, which is determined by the contents of the pulse. The content of the pulse that produces adequate entropy for a sufficiently random CID includes internal randomness and external randomness. The internal randomness, as described in NIST beacon specification~\cite{kelsey_randomness_2019}, is the hashed combination of three independent pseudorandom bit generators. In this case, the three used are:

\begin{enumerate}
    \item OpenSSL's PRNG (accessed via the node.js \code{crypto.randomBytes()} function)
    \item Hardware randomness from YubiHSM2’s hardware security module
    \item A custom written RSAPRNG based on the RSA algorithm with special safe prime selection (see section~\ref{rsaprng} for details)
\end{enumerate}

External randomness comes from the pulse mixin field which includes the hashes of pulses on other chains—specifically, the CURBy-Q pulses and DRAND pulses contain regular verified randomness.

\begin{figure}
    \centering
    \includegraphics[width=0.75\linewidth]{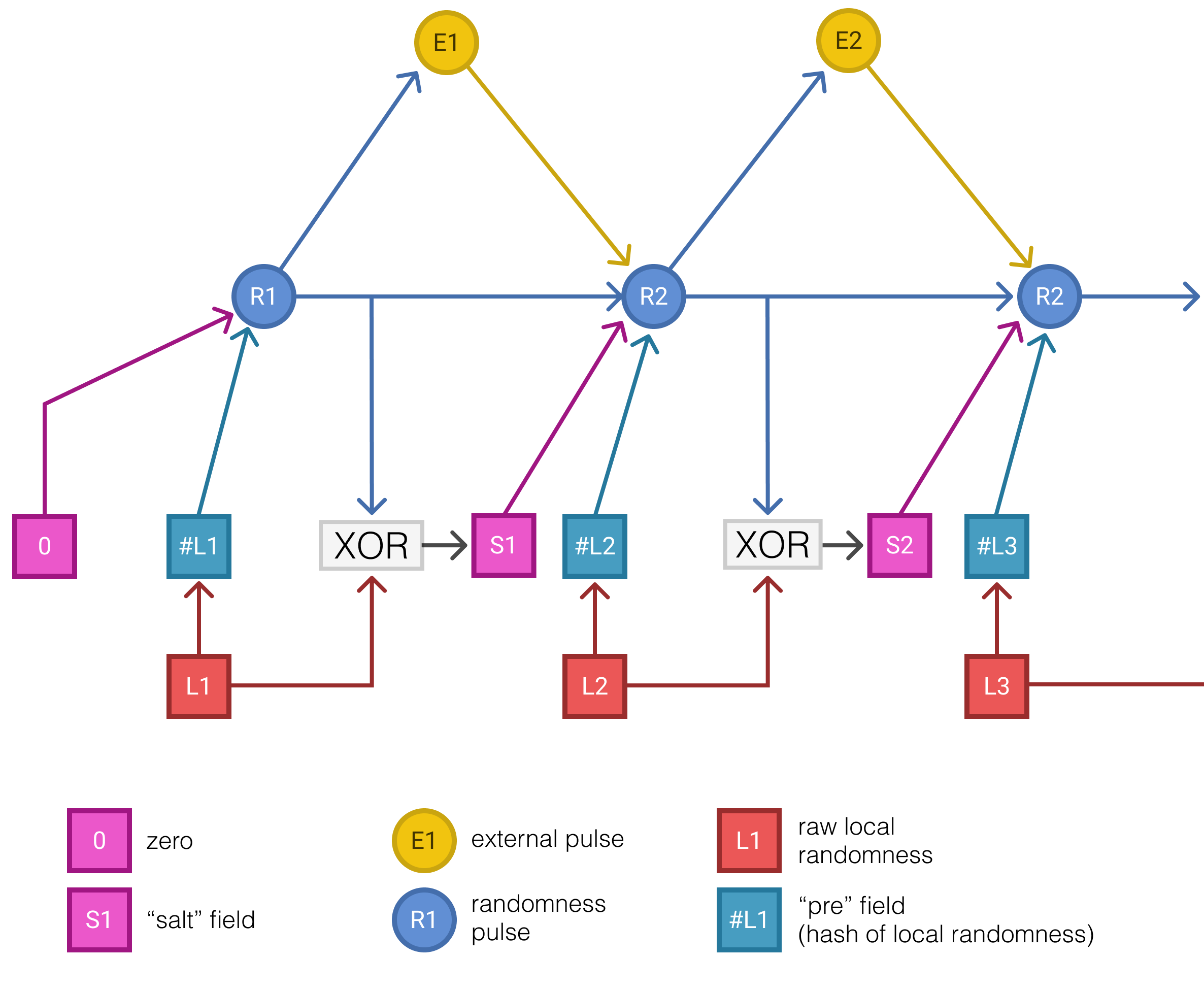}
    \caption{Flow diagram depicting the payload assembly of a PRNG pulse in the NIST-inspired beacon.}
    \label{fig:prngpulse}
\end{figure}

The process for generating a PRNG pulse is described by figure~\ref{fig:prngpulse} and includes many of the bias-mitigation strategies devised in the NIST randomness beacon specification~\cite{kelsey_randomness_2019}. The three local PRNG sources are hashed together creating the raw local randomness. A precommitment (\code{pre} field) is created by again hashing the local randomness as a mechanism to commit to using those bits to create the next pulse’s \code{salt} field. The \code{salt} field begins populated with a 512 bitstring of zeros making the \nth{0} pulse invalid for randomness. Each subsequent pulse’s \code{salt} field is the XOR of the previous pulse’s output hash (the hash portion of the CID) and the raw local randomness. The purpose of all of these steps is to mitigate the opportunity for introducing bias into the output randomness by the process creating the pulses (CU), while also injecting randomness unknowable to the public until the pulse is published. 

When consuming the randomness, it is highly recommended to perform several verification steps on the pulse to ensure the protocol was followed honestly. The CURBy javascript client library performs these verifications.

\subsection{Seam Chain}

The seam chain is the simplest chain. Its sole purpose is to provide a logical separation between the core CURBy chains and third-party chains. Every pulse entwines with its neighboring chains and simply contains a timestamp in the payload.

\subsection{Weather Chain}

The weather chain is a simple proof-of-concept of a third-party using Twine for its own application while still being interoperable with the CURBy network. The result is an enhancement of the integrity of the audit trail since an updated hash-link trail can always be followed from the third-party pulses back to CURBy pulses and vice versa. The weather chain publishes a pulse every 5 minutes with a payload consisting of the raw API output of Boulder weather from \href{http://openweathermap.org}{\url{openweathermap.org}}.

\section{RSAPRNG} \label{rsaprng}

One of the randomness sources for the PRNG chain is a cryptographically secure pseudorandom number generator based on repeatedly encrypting an initial random seed $x_0$ using the RSA~\cite{rivest_method_1978} encryption algorithm
\begin{align}\label{eqn:encrypt3}
x_k&= x_{k-1}^e\modu n , \notag\\
&= x_0^{e^k \modu \lambda(n)} \modu n,
\end{align}
where $n=p q$ is the product of two large randomly chosen primes and the exponent $e$ is coprime to Euler's totient function 
\begin{align*}
\phi(n)=(p-1)(q-1) .
\end{align*}
Unlike public key RSA encryption, the modulus $n$ and exponent $e$ are never revealed.

Repeated application of equation~\eqref{eqn:encrypt3} defines a stream cipher operated in output feedback mode (OFB), and the security of the algorithm is based on the well-established security of the RSA algorithm~\cite{kohno_cryptography_2010}. We generate 512-bit cryptographically secure pseudorandom numbers by repeatedly encrypting the $x_k$ using equation~\eqref{eqn:encrypt3} and concatenating together the least significant bit of sequential values of $x_k$~\cite{menezes_handbook_2018}.

The properties of the generator can be analyzed using elementary number theory~\cite{koshy_elementary_2002, silverman_friendly_2006, koblitz_course_1987}. 
The period of the generator is determined by the multiplicative order function
$\ord_n(a)$ which is defined for all $a\in\mathbb{Z}_n$ coprime to $n$ and is the smallest integer $t$ such that $a^t \mod n = 1$. The order $\ord_n(a)$ divides 
the Carmichael reduced totient function 
 $\lambda(n)$ which is the maximal multiplicative order, i.e. the largest value of $\ord_n(a)$ for all $a$ coprime to $n$. 
For prime powers $\lambda(n)$ is simply related to $\phi(n)$ 
\begin{align}
\lambda(p^\alpha) = 
\left\{ 
\begin{array}{lll}
 &\phi(p^\alpha)=p^{\alpha-1} (p-1), \quad &\mbox{for odd primes},\\ 
 &\phi(2^\alpha)=2^{\alpha-1},\quad &\mbox{for $\alpha=1,2$},\\
 &\frac{1}{2}\phi(2^\alpha)=2^{\alpha-2},\quad &\mbox{for $\alpha>2$}.
\end{array}
\right.
\end{align}
For general value $n$ expressed as the product of its prime factors $n=\prod_j p_j^{\alpha_j}$, $\lambda(n)$ is the least common multiple of the Carmichael totients of the prime power factors:
\begin{align}
\lambda(n) = \lcm(\lambda(p_1^{\alpha_1}),\lambda(p_2^{\alpha_2}),\ldots) .
\end{align}

Since $\ord_n(x_0)$ divides $\lambda(n)$ and $\ord_{\lambda(n)}(e)$ divides $\lambda(\lambda(n))$, we can choose p and q so that both $\lambda(n)$ and $\lambda(\lambda(n))$ have large prime factors.

This can be accomplished by choosing $p$ of the form $p=2 a_1 p_1 +1$ and $p_1=2 a_2 p_2 +1$ where $p_1$ and $p_2$  are large primes, and likewise for $q$. We choose the bit-lengths of $p_1,p_2,q_1,q_2$ consistent with the natural distribution of largest prime factors of randomly chosen integers. 
The largest prime factor of a randomly chosen integer $n$ will be greater than $n^x$ with probability approximately $-\ln x$ for $x>1/2$~\cite{knuth_analysis_1976,riesel_prime_1994}.

For example, the probability that the largest prime factor of a random integer $n$ is greater than $n^{3/4}$ is $\ln(4/3) \simeq 0.29$. 
We can use this property to randomly select $p$ and $q$ with the properties $p_1>p^{3/4}$ and 
and $p_2>p_1^{2/3}>p^{1/2}$; likewise for $q$. 
Choosing $p$ and $q$ using this distribution will assure that the chosen primes $p$ and $q$ are not atypical and provides assurance with high probability that the period is greater than $\sqrt{n}$. For our implementation of equation~\eqref{eqn:encrypt3} we choose $p$ and $q$ to each be of the order of $2^{1536}$.

\bibliography{beacon_paper_references}